\documentclass{aastex}
\usepackage{emulateapj5, epsfig, apjfonts}
\reversemarginpar
\makeatletter
\newenvironment{tablehere}
  {\def\@captype{table}}
  {}
\newenvironment{figurehere}
  {\def\@captype{figure}}
  {}
\makeatother

\newcommand{\percm}{${\rm cm^{-2}}$}
\newcommand{\kms}{${\rm km \ s^{-1}}$}

\newcommand{\lam}{$\lambda$}
\newcommand{\lamlam}{$\lambda \lambda$}
\newcommand{\lya}{Ly$\alpha$}
\newcommand{\lyb}{Ly$\beta$}
\newcommand{\nhi}{\mbox{$N($\ion{H}{1}$)$}}
\newcommand{\nciv}{\mbox{$N($\ion{C}{4}$)$}}
\newcommand{\nov}{\mbox{$N($\ion{O}{5}$)$}}
\newcommand{\ovhi}{\mbox{$\langle$\ion{O}{5}/\ion{H}{1}$\rangle$}}
\newcommand{\ovihi}{\mbox{$\langle$\ion{O}{6}/\ion{H}{1}$\rangle$}}
\newcommand{\oh}{\mbox{$\langle {\rm O} / {\rm H} \rangle$}}

\newcommand{\civhi}{\mbox{$\langle$\ion{C}{4}/\ion{H}{1}$\rangle$}}
\newcommand{\oivov}{\mbox{$\langle$\ion{O}{4}/\ion{O}{5}$\rangle$}}
\newcommand{\xihi}{\mbox{$\langle {\rm X_i}$/\ion{H}{1}$\rangle$}}
\newcommand{\hm}{HM96\nocite{hama96}}
\newcommand{\jnu}{$\rm ergs\ cm^{-2}\ s^{-1}\ Hz^{-1}\ sr^{-1}$}

\lefthead{TELFER ET AL.}
\righthead{EUV ABSORPTION LINES IN LYMAN FOREST ABSORBERS}

\begin{document}
\submitted{Accepted for publication in November 10, 2002 edition of the Astrophysical Journal}
\title{Extreme Ultraviolet Absorption Lines in \lya\ Forest Absorbers and the Oxygen Abundance in the
Intergalactic Medium}
\author{Randal C. Telfer\altaffilmark{1}, Gerard A. Kriss\altaffilmark{1,2},
Wei Zheng\altaffilmark{1}, Arthur F. Davidsen\altaffilmark{1,3}, David Tytler\altaffilmark{4,5}}
\altaffiltext{1}{Center for Astrophysical Sciences, Johns Hopkins University, 
Baltimore, MD, 21218-2686}
\altaffiltext{2}{Space Telescope Science Institute, 3700 San Martin Drive, 
Baltimore, MD, 21218}
\altaffiltext{3}{Deceased 2001 July 19}
\altaffiltext{4}{Center for Astrophysics and Space Science, University of California, San Diego, 
La Jolla, CA, 92093-0424}
\altaffiltext{5}{Visiting Astronomer, W.\ M.\ Keck Observatory, which is a joint facility of the
University of California, the California Institute of Technology, and NASA}

\begin{abstract}
We create stacked composite absorption spectra from {\it Hubble Space Telescope} 
Faint Object Spectrograph
data from four quasi-stellar objects to search for absorption lines in the extreme ultraviolet 
wavelength region associated with \lya\ forest absorbers in the redshift range $1.6 < z < 2.9$. 
We successfully detect \ion{O}{5} \lam 630 in
\lya\ absorbers throughout the $10^{13}$ to $10^{16.2}$ \percm\ column density range. For a sample
of absorbers with $10^{13.2} <$ \nhi\ $< 10^{14.2}$ \percm, corresponding to gas densities ranging
from around the universal mean to overdensities of a few, we measure an \ion{O}{5} \lam 630 
equivalent width of $10.9 \pm 3.7$ m\AA.  We estimate the detection is real with at least 99\% confidence.
We only detect \ion{O}{4} \lam 788, \ion{O}{4} \lam 554, \ion{O}{3} \lam 833, and \ion{He}{1}
\lam 584 in absorbers with \lya\ equivalent widths $\gtrsim 0.6$ \AA, which are likely associated
with traditional metal-line systems.  We find no evidence in any subsamples for absorption from
\ion{N}{4} \lam 765, \ion{Ne}{5} \lam 568, \ion{Ne}{6} \lam 559, \ion{Ne}{8} \lamlam 770, 780, or
\ion{Mg}{10} \lamlam 610, 625.
The measured equivalent widths of \ion{O}{5} suggest values of \ovhi\ in the range $-1.7$ to $-0.6$ for 
$10^{13.2} <$ \nhi $\lesssim 10^{15}$ \percm.  
The lack of detectable \ion{O}{4} absorption except in the strongest absorption systems 
suggests a hard ionizing 
background similar to the standard Haardt \& Madau spectrum.  Using photoionization
models, we estimate that the oxygen abundance in the intergalactic medium with respect to the 
solar value is $[{\rm O}/{\rm H}] \approx -2.2$ to $-1.3$.
Comparing to studies of \ion{C}{4}, we estimate
$[{\rm O}/{\rm C}]\approx 0.3$ to 1.2.  
The overabundance of oxygen relative to carbon agrees with other low-metallicity 
abundance measurements and suggests enrichment of the intergalactic medium by Type II supernovae.
\end{abstract}


\section{INTRODUCTION\label{sec:intro}}

Early studies of the intergalactic medium (IGM) identified two separate groups of absorbers 
\citep[e.g.,][]{sybt80}:  (1) ``metal-line systems'' which have observable metal lines and the 
strongest \lya\ absorption, typically with a \lya\ rest-frame equivalent width $\gtrsim$ 1 \AA,
and (2) weaker absorbers having no observable metal lines and comprising what is generally 
thought of as the ``\lya\ forest''.  However, it was suspected that this distinction was due at 
least partly to observational limitations in the observability of weak metal lines 
\citep{tytl87}, a suspicion that has been verified as technological and computational 
advances have allowed astronomers to detect metal absorption to much lower column densities.
Constraining the abundances of metal ions in the IGM has become an active area of research,
since this information can in principle be used to determine the chemical abundances in the 
IGM as well as shape and strength of the typical metagalactic ionizing background radiation
from stars and quasi-stellar objects (QSOs).  Of particular interest is the low column density regime 
(\nhi $\lesssim 10^{14.5}$ \percm), since these systems are presumably far away from 
local sources of metals and ionizing photons and thus offer the best insights into
enrichment mechanisms and the ionizing background.

The production of composite or stacked absorption spectra of many \lya\ absorbers is an obvious way
to increase observational sensitivity.  \citet*{nhp83} used the composite technique to
claim a detection of \ion{O}{6} \lamlam 1032, 1038 absorption.  The reality of this detection
was challenged in a similar study by \citet{wcwb+89}, but \ion{O}{6} in the IGM was detected
unambiguously by \citet{lusa93} by creating a composite absorption spectrum of \ion{C}{4}
absorbers.  Composite spectra were used to detect \ion{C}{4} absorption in systems
with \nhi $\gtrsim 10^{14}$ \percm\ at high redshift ($z > 1.7$) by
\citet{lu91}, but \citet{tyfa94} failed to detect \ion{C}{4} associated with weaker lines.
In a study similar to \citet{lu91} but at $z < 0.8$, \citet{baty98} found an order of magnitude
stronger \ion{C}{4} absorption than \citet{lu91}.

A more precise way to increase sensitivity is to build bigger telescopes and better instruments
in order to acquire higher signal-to-noise ratio (S/N), higher resolution data.   With the HIRES 
instrument on the 10m Keck~I telescope, individual \ion{C}{4} features can be detected down to a 
column density of  \nciv $\sim 10^{12}$ \percm.
The majority of \lya\ absorbers with \nhi $> 3 \times 10^{14}$ \percm\  exhibit such 
absorption \citep{tfbc+95,cskh95,soco96}.
Besides allowing for the direct detection of weaker lines, the existence of such high quality data 
has opened the door to an additional detection method introduced by \citet{coso98}, that of
comparing the individual pixel optical depths of \lya\ to the expected position of corresponding 
metal lines.  \citet{coso98} and \citet{essp00} use this technique to detect \ion{C}{4} 
absorption down to an \ion{H}{1} optical depth of $\tau_{\rm HI} \lesssim 1$.  \citet{srsk00} have
used the optical depth technique to detect \ion{O}{6} to optical depths down to
$\tau_{\rm HI} \sim 0.1$.

Studies have concentrated on \ion{C}{4} not only due to the fact that it is expected to be one
of the strongest features associated with the IGM (\citealt*{rhs97}; \citealt{hhkw98}), 
but also largely because
of its near ideal location in wavelength space.  With a wavelength of 1549 \AA, this absorption 
feature is near enough to \lya\ \lam 1216 that both features can easily be acquired with the same
observations, but far enough above \lya\ that the feature will be longward of the \lya\ forest
for a significant span in redshift.  Although \ion{C}{4} is a good tracer
of metals in the IGM, in standard photoionization models 
the \ion{O}{6}  \lamlam 1032, 1038 \AA\ doublet is expected to be a 
stronger feature at \ion{H}{1} column densities $\lesssim 10^{15}$ \percm\ \citep{hhkw98}.
Unfortunately, although the wavelength of \ion{C}{4} makes
it very appealing for absorption studies, the \ion{O}{6} \lamlam 1032, 1038 \AA\ doublet could
hardly reside in a worse portion of the spectrum.  The region of \ion{O}{6} absorption is muddled
not only by \lya\ absorption but by higher order Lyman lines as well, making detections much more
difficult.  Fortunately, \citet{srsk00} have overcome this difficulty to detect \ion{O}{6} in
the diffuse IGM, although they make no quantitative claims as to the implied amount of \ion{O}{6}
in the IGM.

The extreme ultraviolet (EUV) absorption region has been largely ignored, not because this
region contains no useful information, but because the best available data in this region,
that from the {\it Hubble Space Telescope} ({\it HST}) Faint Object Spectrograph (FOS), is
inferior in resolution and S/N to that used at longer wavelengths.  However,
some of the strongest expected absorption features occur in this region, in particular
\ion{O}{5} \lam 630 and \ion{O}{4} \lam 788 \citep*{vtb94}.  The \ion{O}{5} feature
is especially worthy of interest because its large oscillator strength ($f = 0.514$,
greater than \ion{H}{1} \lya) should make it competitive in strength to the \ion{O}{6}
feature even at column densities down to $\sim 10^{13}$ \percm.  
Despite the fact that
the FOS has rather low resolution by modern standards ($R \approx 1300$), 
searching for \ion{O}{5} rather than
\ion{O}{6} has the advantage that the features are in a spectral region where the 
\lya\ forest is less dense owing to the strong evolution of the density of lines with redshift.
The detection of oxygen lines in the IGM can be used to infer the
oxygen abundance, which is expected to be overabundant relative to carbon with respect to the solar
ratio if the early universe is enriched by Type II supernovae.  Strong evidence for such abundance 
patterns are seen at high redshift in Lyman-limit
systems \citep{rvhe+92,krth+99}, damped \lya\ absorbers (\citealt*{plh95}; \citealt{lsbc+96}), and \ion{C}{4}
absorbers \citep{dhhk+98,song98}.  Here we extend this work to the lower column density regime by
carrying out a statistical search for absorption in the EUV using the presently available data.

In \S\ref{sec:data} we discuss the data and the spectral selection criteria that we use for
this study.  We begin \S\ref{sec:analysis} by discussing the characterization of our \lya\
absorber sample from optical data.  This is followed by a presentation of the details of the 
technique that we use to generate the stacked absorption spectra and the method for 
measuring the equivalent width of the resulting absorption features.  In \S\ref{sec:results}
we describe the results of our search for various samples selected by the strength of \lya\ 
absorption, particularly the detections of \ion{O}{5}.  We consider the implications of our
results in \S\ref{sec:discuss} by comparing the measurements and limits on the equivalent widths
of the absorption features to simulated data to constrain the
abundance ratios of the relevant ions to \ion{H}{1}.  Using photoionization models, we then infer
the abundance of oxygen in IGM.  We provide a brief summary of our results in \S\ref{sec:summary}.

\section{DATA\label{sec:data}}
We have searched the {\it HST} archives for QSO spectra suitable for detecting 
EUV absorption features in the \lya\ forest, with the primary goal being the detection
of \ion{O}{5} \lam 630.  To include an object in this study, we require that:
\begin{enumerate}
\item{Data are available shortward of 630 \AA\ in the QSO emission frame, making the 
\ion{O}{5} feature accessible.}
\item{The data are obtained with the FOS high-resolution gratings.}
\item{The S/N in the region of the \ion{O}{5} absorption is high
enough that the rms deviation of the pixels is dominated by absorption features rather than
Poisson noise.  For the typical density of lines in the applicable redshift range 
($1.6 < z < 2.9$), the necessary S/N is $\sim 5$ per pixel.  At this S/N, we
can make a fairly reliable determination of the continuum level, which is necessary for
our study.}
\item{We have access to high-quality optical spectra for determining the redshift
and strength of the \lya\ forest lines.}
\end{enumerate}
We have found four QSOs that match these criteria.  The FOS data for these QSOs were 
reduced in 2001 July with the best available reference files.
In Table~\ref{ta:fosdata} we
list these objects, their redshifts, the total exposure time of the FOS data with each
grating, and the dates the data were obtained. The zero-point shift for the wavelength scale
given in of the last column will be discussed in \S\ref{sec:wave}.  All of the data
were obtained with the red Digicon detector on the FOS.  A G190H spectrum of HS~1700+6416
was obtained with the blue Digicon on 1997 May 29.  However, the data are of somewhat
lower resolution and for consistency we use only the data from the red Digicon.

\begin{table*}
\caption{FOS Data\label{ta:fosdata}}
\centerline{
\begin{tabular}{lccclc} \hline \hline
Object	&	$z$	&
Grating	&	Exposure (s) &
Date	&	Shift (\AA)\\
\hline
HS~1103+6416	& 2.190 & G270H & 5336 & 1996 Oct 31 & $+0.10$\\
		& 	& G190H & 8628 & 1996 Oct 31 & $+0.59$\\
HE~1122--1649	& 2.400 & G270H & 780  & 1996 Jul 18 & $+0.30$\\
		&	& G190H & 5620 & 1996 Jul 18 & $+1.25$\\
HS~1700+6416	& 2.743	& G270H & 9000 & 1992 Dec 13 & $0.00$\\
		&	& G190H & 16120 & 1992 Dec 12--13 & $+0.85$\\
HE~2347--4342	& 2.885 & G270H & 1320 & 1996 Jun 7  & $+0.80$\\
		&	& G190H & 7930 & 1996 Jun 7  & $+1.05$\\
\hline
\end{tabular}}
\end{table*}

For all four objects we have access to Keck HIRES spectra for examining the \ion{H}{1}
\lya\ forest.  The spectrum for HE~2347--4342 is courtesy of A.\ Songaila and is discussed in
\citet{song98}.  Two of the other three spectra are discussed elsewhere,
and we point the reader to the corresponding articles for details:
\citet{krth+99} for HS~1103+6416 (1997 April 9 data only) and \citet{drtb+00} for HE~1122--1649.
The spectrum of HS~1700+6416 consists of a 4300 second exposure on 1995 May 10.  This spectrum covers
wavelengths from 3727 \AA\ to 5523 \AA\ with a typical S/N of $\sim 50$ per resolution element.

\section{ANALYSIS\label{sec:analysis}}

\subsection{\lya\ Forest Data\label{sec:forest}}
The Keck HIRES data are normalized to the continuum by fitting a spline by hand,
placing nodes at what appear to be unabsorbed portions of the spectra.  We then
fit the spectra in sections of $\sim 50$ \AA\ down to the redshifted position of \lyb.
We use the IRAF task {\it specfit} to fit the individual absorption features
with Gaussians in optical depth:
\begin{equation}
\tau = \tau_0 \ \exp \left[{-\left( \frac{\lambda - \lambda_0}{\lambda_0 b / c}\right)^2}\right].
\end{equation}  
Each
absorption feature is thus described by three free parameters:  the central wavelength $\lambda_0$,
the optical depth at line center $\tau_0$, and the Doppler width $b$.  In a few cases where
the blending is severe, the $b$-parameter is fixed at 30 \kms, a typical value
for \lya\ forest lines \citep{rccf+92, hkcs+95}, to prevent $b$ from taking on an unreasonable
value.  We fit
all lines with peak optical depths $\gtrsim 0.1$, although lines weaker than this
were sometimes fit to deblend stronger lines.

We impose several selection criteria on our list of \lya\ forest lines:
\begin{enumerate}
\item{As is frequently done in \lya\ forest studies, we ignore lines shortward of 
\lyb\ in the rest frame of the QSO to avoid confusion of \lya\ lines with higher-order 
lines.}
\item{We discard lines within 5000 \kms\ of the QSO rest frame in order to avoid
contamination from the proximity effect \citep[see e.g.][]{sbdk00}.  This should also eliminate
many associated absorbers, which often occur near the QSO in velocity space.  However, some
known associated absorbers are displaced substantially more than 5000 \kms\ blueward of the QSO
redshift (\citealt{jhks+96}; \citealt*{hbj97}).  We assume that such systems are rare and not a 
significant source of contamination in our sample.}
\item{We do not use lines that correspond to known Lyman-limit systems.  These 
high-column-density systems have been well-studied
on an individual basis \citep{vore95}.  Since we have data at the Lyman limit for all
candidate lines by virtue of our choice of sample, it is easy to identify Lyman
limit features with $\tau \gtrsim 0.1$, corresponding to \ion{H}{1} 
column densities $\gtrsim 10^{16.2}$ \percm.}
\item{We eliminate lines that are identified as likely metal lines associated with
Lyman-limit or metal-line absorption systems.}
\item{We ignore features in wavelength regions that are contaminated by strong 
clusters of such metal lines.}
\item{We dismiss lines with $b < 14$ \kms\ as probable unidentified metal lines.
Detailed studies of the distribution of $b$ \citep{hkcs+95,kity97} show that
there is a sharp cutoff at around 20 \kms\ and virtually no lines with 
$b < 14$ \kms.}
\item{We also exclude lines with $b > 80$ \kms.  Although a few of these lines are
probably real, they are certainly atypical.  Some are likely due to blending.}
\end{enumerate}

It is straightforward to use the best-fit Gaussian parameters to calculate the
equivalent width, redshift, and column density of \ion{H}{1}.  To illustrate that
we have extracted a reasonable distribution of lines, we plot the number of lines
in our sample as a function of column density in Figure~\ref{fig:cddf}.  The median 
redshift of the lines in our sample is $z=2.35$.
We have placed
all of the lines between $10^{14.2}$ and $10^{16.2}$ \percm\ into a single bin since
the column densities of these strongly saturated lines are poorly determined.  In principle,
we could use higher-order Lyman lines to measure the column density.  Unfortunately, for most
of the \lya\ features in our sample, the corresponding \lyb\ feature is either in the low wavelength
portion of the Keck data, where the S/N is low, or at wavelengths in between the Keck and FOS spectral
coverage where we have no data.

\begin{figurehere}
\centerline{\psfig{file=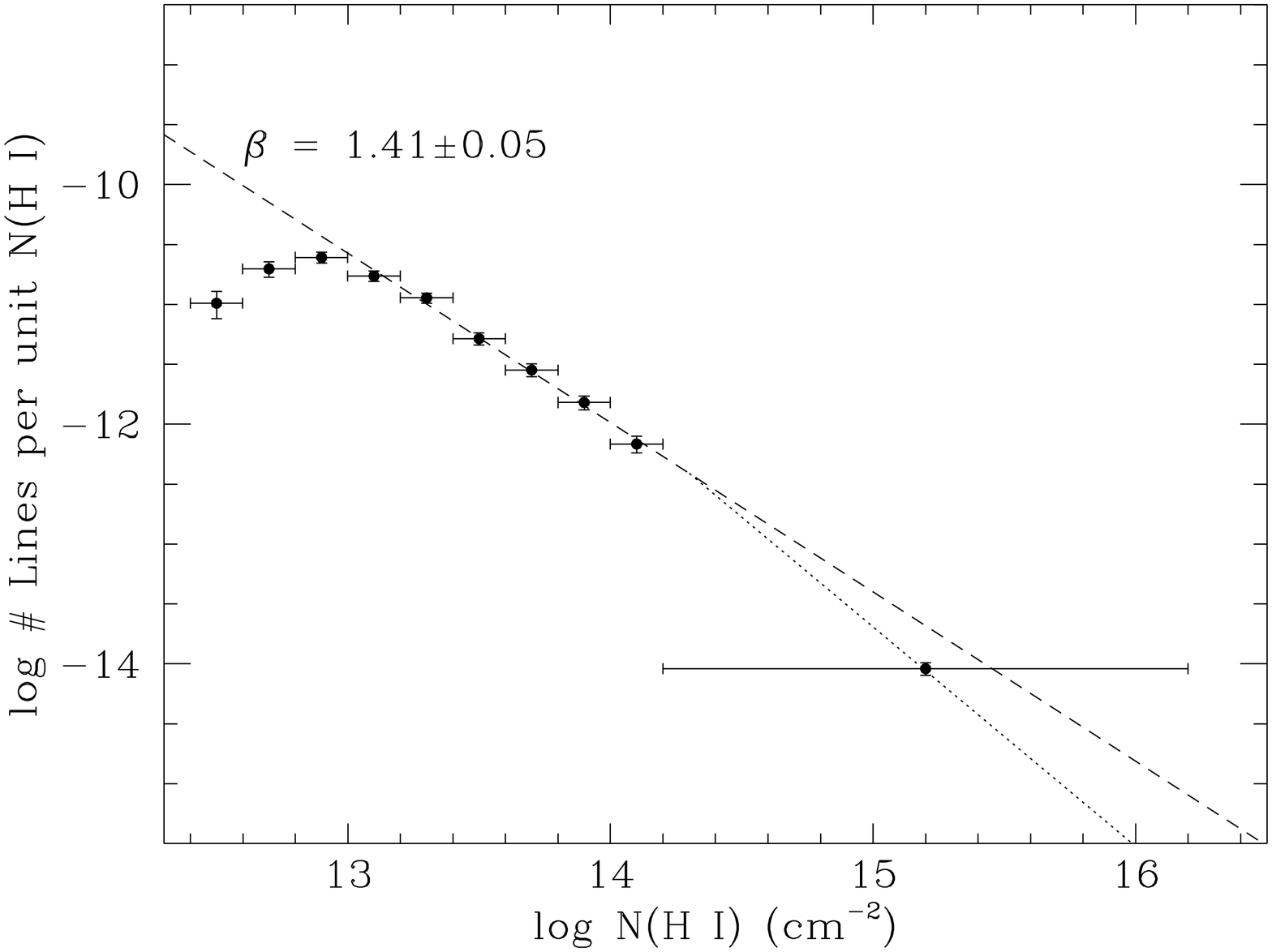, angle=-90, width=9cm}}
\caption{Distribution of the number of lines in our sample with column density.  The 
dashed line is our best fit to the distribution between $10^{13}$ and
$10^{14.2}$ \percm\ with $\beta = 1.41 \pm 0.05$. The dotted line represents
$\beta = 1.83$ connecting to our best fit at $10^{14.3}$ \percm.\label{fig:cddf}}
\end{figurehere}
\vspace{0.4 cm}

The ordinate in Figure~\ref{fig:cddf} is proportional to the column density distribution
function, often called $f(N)$ in the literature.  As is standard practice, we fit the
distribution between $10^{13.0}$ and $10^{14.2}$ \percm\ with a power law, 
$f(N) \propto N^{-\beta}$.  Our best-fit power-law index is $\beta = 1.41 \pm 0.05$,
plotted as the dotted line in Figure~\ref{fig:cddf},
in agreement with \citet{khcs97} for this range of column densities at $z\approx 2.3$.
The fit appears to be good down to $10^{13}$ \percm, which argues that we are 
essentially complete to this column density.
The column density distribution function is known to steepen above $10^{14.3}$ \percm\
\citep{khcs97}.  We adopt $\beta = 1.83$ as found by \citet{pwrc+93} to be a good fit
between $\sim 10^{14}$ and $10^{16}$ \percm.  We connect this distribution to our best
fit at $10^{14.3}$ \percm\ and plot it as the dotted line in Figure~\ref{fig:cddf}.
Clearly this agrees well with our high-column-density point.  When considering
the distribution of column densities for \nhi $> 10^{14.3}$ \percm\, which is important 
for our data simulations in
\S\ref{sec:abund}, we therefore adopt $\beta = 1.83$ as 
representative of our data.

\subsection{Combination Technique\label{sec:combine}}
The process of creating composite absorption spectra is conceptually straightforward.
First we combine the G190H and G270H spectra into a single spectrum so that the 
wavelengths in the overlap region do not receive double weighting.  To facilitate this, we
resample the individual spectra into 0.1 \AA\ bins in the observer frame -- a factor of several smaller 
than the 
original pixels so as not to alter the character of the data.  Then we normalize 
the FOS spectra to the continuum using a cubic spline in the
same way as we did for the optical data.  Instrumental
artifacts and a damped \lya\ absorption feature in one object are masked out.  
We also remove the noisy portion at the short wavelength
end of the G190H data, below $\sim 1650$ \AA, where the sensitivity of the red Digicon
detector becomes low and the S/N drops per original FOS pixel drops below 5.

For each \lya\ forest feature in the
particular sample, the combined FOS spectrum of that object is shifted into the rest frame of
the absorber by dividing the wavelength scale by $1+z_{\rm abs}$.  These shifted spectra are
then rebinned into a common wavelength scale in the absorber frame for coadding.  For convenience we choose
a bin size of 0.1 \AA,  which corresponds to 48 \kms\ in the region of 
\ion{O}{5} \lam 630, slightly smaller than the typical pixels in the original data.

For each pixel in the proposed composite spectrum, we then have a distribution of pixel fluxes
from the individual shifted spectra.
A crucial question remains -- what statistic of this distribution do we use as the composite flux
in order to 
optimize the sensitivity of our measurements?  The answer depends strongly on the character of 
the noise, which in our study is due largely to random absorption features.  
If we were dealing with high S/N data and the continuum in the spectral region of
interest were not polluted by random absorption lines, as is typical in searches for
weak \ion{C}{4} absorption, then the noise would be Gaussian distributed and a mean or median 
combination method would do nicely to minimize the noise in the result.  However, when the 
features of interest are strongly blended into the \lya\ forest the 
noise is distinctly non-Gaussian.

On the top of Figure~\ref{fig:pixhist} we plot as a solid line the histogram of the pixel
fluxes of the FOS spectra of our four QSOs, after the G190H and G270H data have been binned to
0.1 \AA\ and combined.  Clearly the pixels have a very asymmetric distribution in flux with a long
tail to low flux due to the many absorption lines.  We expect that at any random wavelength,
where there is no feature, the contribution to the composite from the individual shifted
spectra will consist of a random sampling from this distribution.
We would like to maximize the S/N in these random samplings, which for our composite spectra
corresponds to the S/N in the continuum.
We can take any general percentile of the data by linearly interpolating
the cumulative distribution, where the lowest flux at a given pixel is defined as the zeroth 
percentile and the highest flux as the 100th.  The simplest and most common example is the 
median or 50th percentile, but we will show that we can increase the S/N significantly by 
making a different choice.

\begin{figurehere}
\centerline{\psfig{file=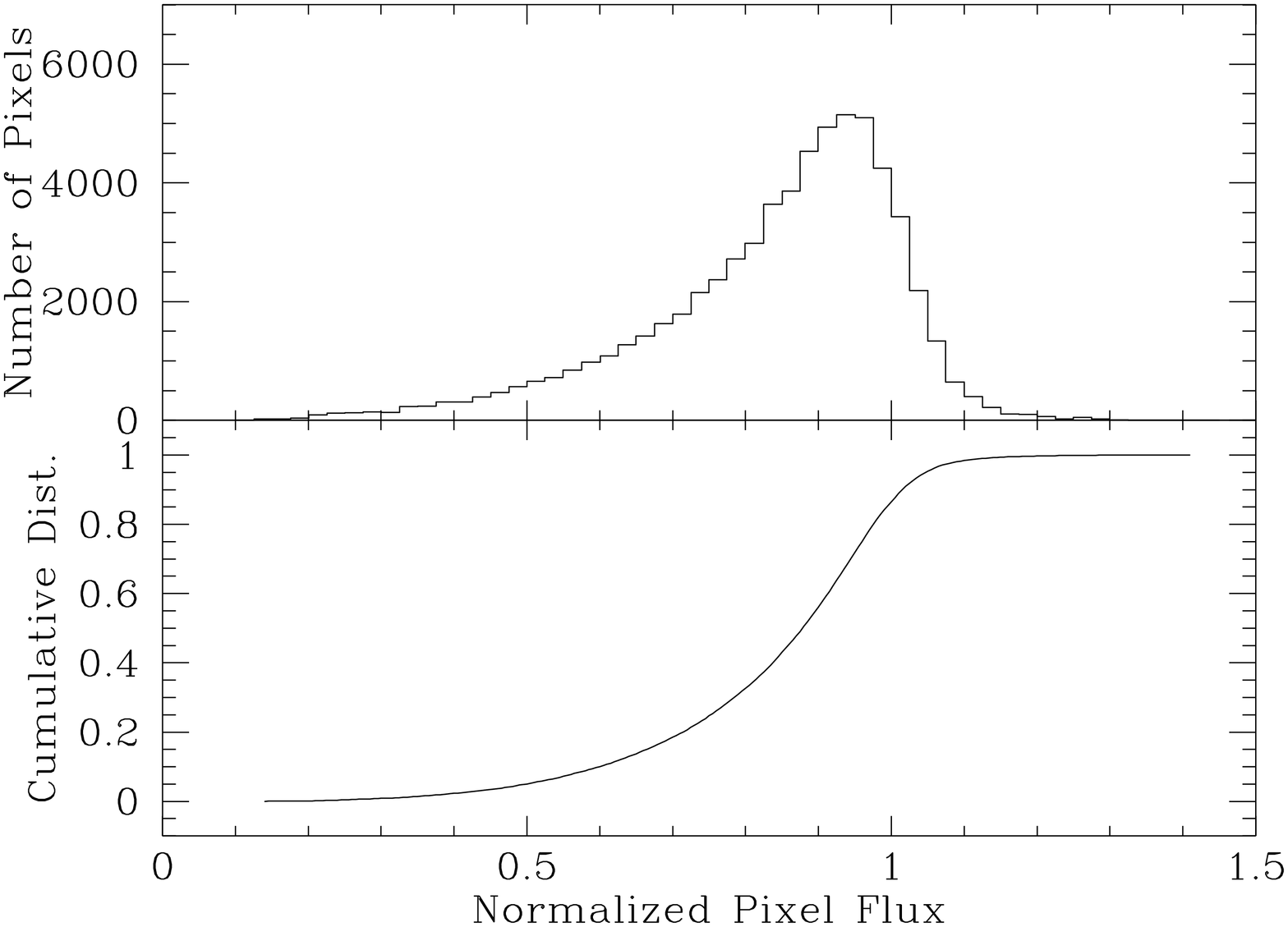, angle=-90, width=9cm}}
\caption{{\it Top}:  Distribution of individual pixel fluxes in the FOS data of the four
QSOs, after the spectra have been normalized and resampled into 0.1 \AA\ bins.
{\it Bottom}:  The same distribution displayed as a
cumulative distribution.\label{fig:pixhist}}
\end{figurehere}
\vspace{0.4cm}

The optimal percentile should be near the
peak of the distribution, which one can see from the cumulative distribution in 
Figure~\ref{fig:pixhist} is around the
80th percentile.  We perform Monte Carlo simulations to show explicitly that this is the case.
We take 100000 random samplings of 100 points from the flux distribution, each time calculating the $n$th
percentile of the sample.  Figure~\ref{fig:snsim} shows the result of the simulations
as a function of percentile, where the S/N is defined as the average value of the $n$th percentile of
the sample divided by the rms deviation.
Indeed, we see that the S/N peaks where the flux distribution
peaks, with the optimal choice being the 79th percentile.  An earlier version of these simulations led us
to conclude that the 78th percentile was optimal which was used in the subsequent analysis.  Since 
Figure~\ref{fig:snsim} shows that the difference is negligible (a 0.3\% difference in S/N), 
we use 78th percentile spectra for
this work.  The 78th percentile yields a S/N nearly 40\% 
greater than the median in these simulations.  The mean yields a S/N of 49.1, similar to the median.
The optimal percentile is not a general result, but rather is peculiar to the S/N and absorption line
density of our data.
This discussion gives our method a solid
statistical basis and points out clearly the importance of carefully considering the noise 
distribution of a particular dataset in order to maximize the resulting S/N in the composite.  
In \S\ref{sec:results} we will demonstrate empirically that taking the 78th percentile of 
the distribution is indeed essentially optimal for our data.

Since we have selected our QSO sample to search for \ion{O}{5} \lam 630, we have FOS data at this
wavelength for virtually all of the \lya\ absorbers on our sample, the only exceptions being those few
for which the wavelength of the \ion{O}{5} absorption happens to lie in a masked region.
Overall, we have FOS data for over 90\%
of the absorbers for rest-frame wavelengths between 560 and 890 \AA.  At wavelengths longer that
$\sim 900$ \AA, the S/N in the composites becomes very low due to the high density of Lyman lines.
As a result, we cannot place any interesting constraints on metal lines at these wavelengths.
There are some interesting lines shortward of 560 \AA, including \ion{O}{4} \lamlam 553, 554 and
\ion{Ne}{6} \lam 559.  Although we could use this \ion{O}{4} doublet to constrain the amount
of \ion{O}{4}, we concentrate on \ion{O}{4} \lam 788 because it provides comparable sensitivity and
we have data for more of the absorbers.  For all other ions that we consider there is 
a single obvious choice for which line to use.

\begin{figurehere}
\centerline{\psfig{file=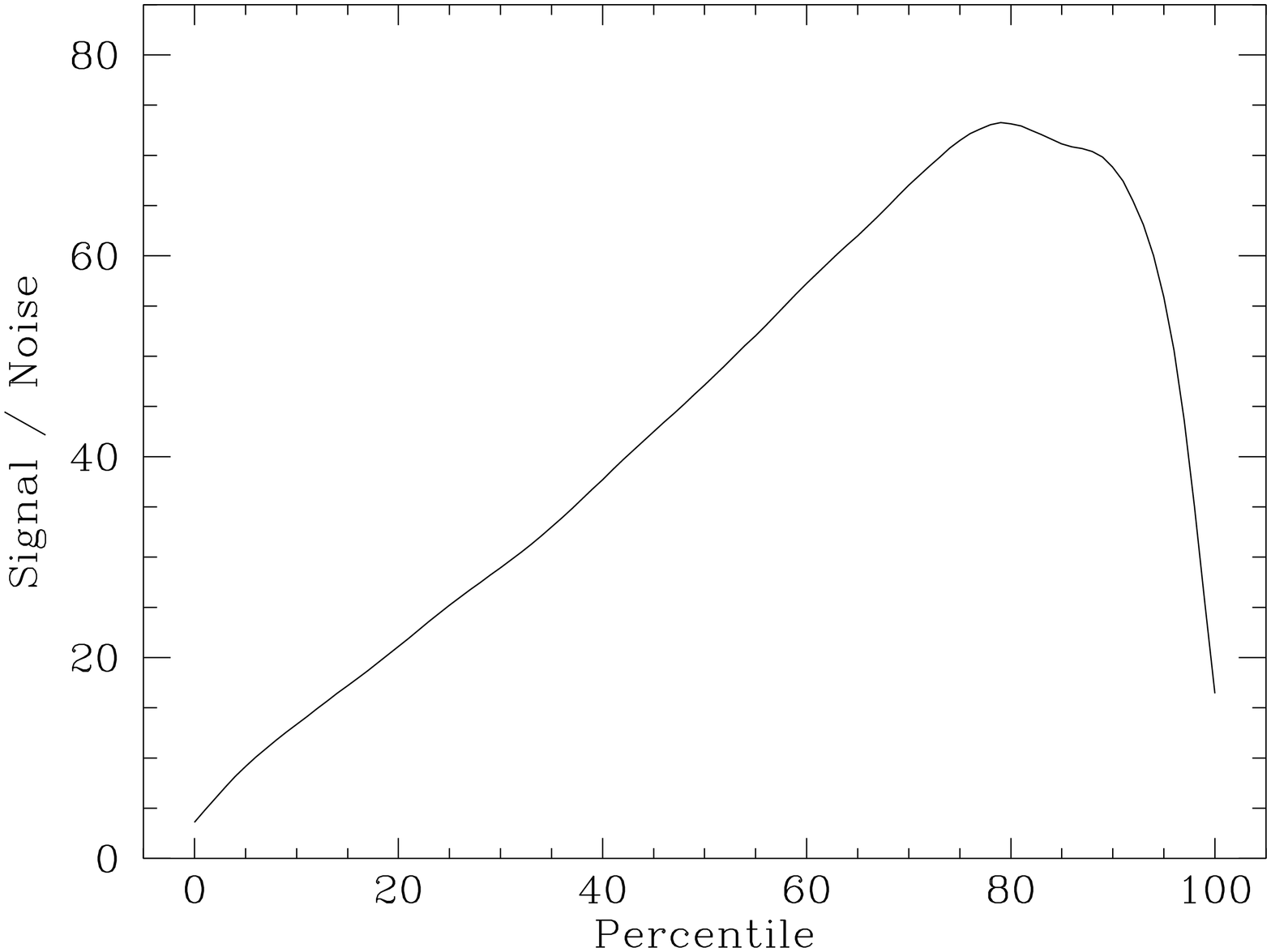, angle=-90, width=9cm}}
\caption{Average S/N of 100000 random samplings of 100 points from
the distribution of all the pixel fluxes from the FOS spectra shown in Figure~\ref{fig:pixhist}, 
as a function of
the percentile taken for each sample.\label{fig:snsim}}
\end{figurehere}
\vspace{0.4cm}

\subsection{Wavelength Calibration\label{sec:wave}}
There is a well-documented problem with the absolute wavelength scale of the FOS
\citep{rkk98}.  A solution to this problem has been found for data from the blue
detector, which has a systematic shift over time, but a fix for the red detector
data has not yet been established.  We determine this empirically for our data.  First, we
find the relative shift between the
G190H and G270H data for each QSO.  Using data in the region of wavelength overlap from
around 2220--2310 \AA, we calculate, to the nearest 0.01 \AA, the wavelength shift
which minimizes the rms deviation of the pixels. We then shift the G190H data by the 
required amount and combine the two into a single spectrum.  
This method works well and gives
shifts consistent with what we estimate by measuring individual features in
the overlap region.  

The overall shift of the combined G190H / G270H data on an absolute scale is
more difficult to determine.  Given the S/N of the data, Galactic absorption lines
are not readily identifiable for HE~1122--1649 and HE~2347--4342, and only a few for 
HS~1103+6416 and HS~1700+6416, so in general using
the Galactic lines is not a reliable way to set the wavelength scale.  To 
estimate the overall shift, we create composite absorption spectra of the individual
objects using only \lya\ forest lines with \lya\ equivalent widths $> 0.5$ \AA.  The lower
limit was chosen so that a well-defined \ion{O}{5} line could be seen for each object.  In each case, 
a strong absorption feature occurs near the expected position of \ion{O}{5} \lam 630, but
slightly shifted by up to $\sim 100$ \kms.  We shift the FOS spectra by the amounts listed in
Table~\ref{ta:fosdata} to center the \ion{O}{5} feature as near as possible to
629.73 \AA.  The shifts listed for the 
G190H data include both the overall shift
and that necessary for agreement with the G270H data. The required shifts are modest and
well within the errors of the FOS red detector.  We use these shifted spectra
for all of our final results.  To check the accuracy of the applied shifts, 
we generate a composite using absorbers with \lya\ equivalent widths $> 0.6$ \AA\ for each object.
The larger lower limit makes some lower-ionization lines more visible,
as discussed in \S\ref{sec:res_h}.
For each object, we have measured the
positions of other lines for which we have data and that are strong enough to make an 
accurate determination of the wavelength centroids.  For this purpose we use lines from the following
list:  Ly$\beta$, Ly$\gamma$, Ly$\delta$, \ion{He}{1} \lam 584, \ion{O}{4} \lam 554, and
\ion{O}{4} \lam 788.  We find that the average shift of the lines for each object is $< 40$ \kms.
Thus, the residual uncertainty in the wavelength scale 
is significantly less than the instrumental resolution and should have little effect on our results.

One may argue that by applying these shifts we could be artificially enhancing the \ion{O}{5} feature, since
we have used \ion{O}{5} to set the wavelength scale.
This may be so, but we point out that such an argument would apply only to \ion{O}{5} and only
for absorbers with a \lya\ equivalent width $>0.5$ \AA.  Our important results are derived almost 
entirely from weaker lines, and there is very little overlap between the relatively weak
absorbers we analyze in detail and the strong systems we use to set the wavelength scale.

\subsection{Measuring the Equivalent Width\label{sec:measew}}
To measure the equivalent widths of the composite absorption features we use
the profile weighting method as described by \citet{lu91}.  Since this is a crucial
part of our analysis we describe it briefly here.  For an unresolved absorption feature, if 
we know the instrumental
line profile and the central wavelength of the feature, then in principle
every pixel gives an estimate of the absorption equivalent width.  To get an
accurate measurement of the equivalent width, we should take a weighted mean of 
the individual pixel estimates of the equivalent width.  If the error is constant
as a function of wavelength, which is roughly true for our data, the optimal weighting
is the square of the instrumental profile.  The equivalent width is then calculated
as
\begin{equation}
{\rm EW} = \frac{\sum (1-f_i / c_i) p_i}{\sum p_i ^2},
\end{equation}
where $f_i$ is the flux, $c_i$ is the continuum level, and $p_i$ is the normalized
instrumental profile.  For the instrumental profile we use a Gaussian with a
full-width at half maximum derived from the composite \lyb\ features, which 
we find to be typically around 290 \kms.  For the continuum determination we use a simple linear
regression fit to the flux between $\pm 2500$ \kms, eliminating the region between
$\pm 500$ \kms\ where the feature lies.

\begin{figure*}
\centerline{\psfig{file=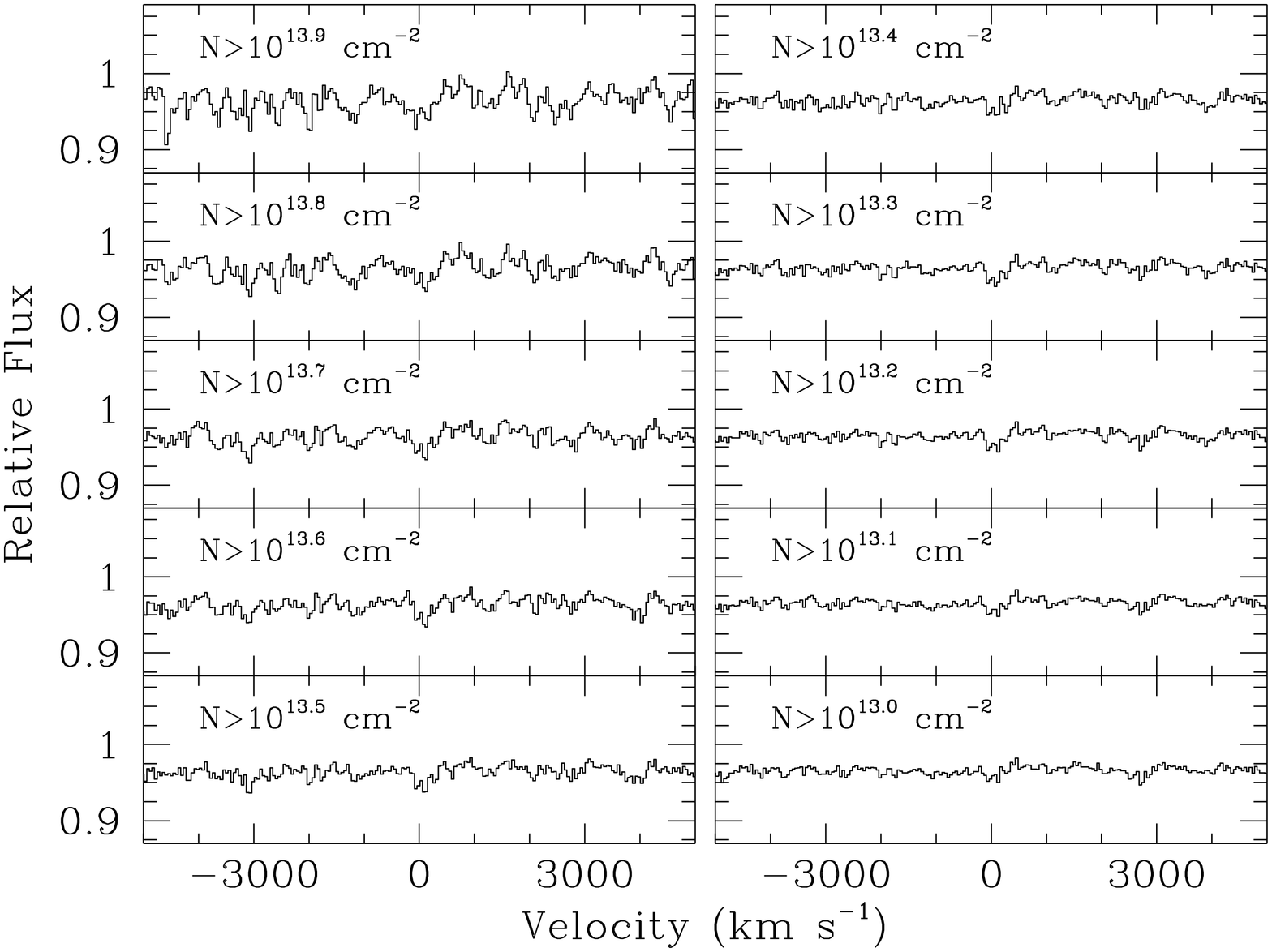,angle=-90,width=14cm}}
\caption{Composite spectra of \lya\ forest features with 
\nhi $< 10^{14.2}$ \percm\ for various minimum \ion{H}{1} column densities, plotted in
the velocity space of \ion{O}{5} \lam 630.\label{fig:lspec}}
\end{figure*}

This weighting technique yields accurate results by relying on the fact that every individual
metal feature associated with the sample has not only the same observed profile shape, but also, 
for any given ion, the same wavelength centroid in the reference frame of the absorber. 
For our sample, we determine the redshift of the absorption systems from the \lya\ absorption lines.
As discussed in \S\ref{sec:wave}, the mean shift of absorption lines with respect to this reference frame
due to uncertainties in the FOS wavelength scale is $< 40$ \kms\ for all four lines of sight.  However, what
are the uncertainties for individual absorbers?  \citet{elpc+99} find an rms dispersion in the velocity
offset between \ion{C}{4} and \ion{H}{1} absorption of 27 \kms, while \citet{essp00} find a dispersion of
17 \kms.  The authors point out that these offsets can have a dramatic effect on the results of the 
stacking technique when applied to high-resolution data, such as that from the Keck HIRES.  However, 
at the FOS resolution, typical offsets of 17--27 \kms\ should have little effect on the composites.

\section{RESULTS\label{sec:results}}

We divide our analysis into three groups based on the \lya\ feature strength.  
One group consists
of the \lya\ lines with \nhi $< 10^{14.2}$ \percm.  These lines are unsaturated and 
therefore we have accurately measured column densities, so the analysis requires fewer
assumptions than for saturated lines.  This is the most important group because it
samples absorbers for which little information is available about metal content.
For stronger lines, we cannot reliably measure the column density of \ion{H}{1}, so
we must instead discriminate absorbers by \lya\ equivalent width.  We divide the stronger
lines into two groups -- an intermediate sample with \lya\ equivalent widths $\lesssim 0.6$ \AA,
and a sample with \lya\ equivalent widths $\gtrsim 0.6$ \AA.  Although the exact division
is somewhat arbitrary, this strongest group consists of lines with \ion{H}{1} column densities
$\gtrsim 10^{15}$ \percm, which are generally associated with metal-line systems.  Each group
thus spans roughly an order of magnitude in column density.  We will discuss the
low-column-density sample first, since these weak lines are useful for assessing the success
of our composite technique.

\subsection{Low Column Densities: \nhi $< 10^{14.2}$ \percm \label{sec:res_l}}
It is not obvious a priori at what \ion{H}{1} column density we should place the lower limit 
of our sample to achieve the most secure detection, and in fact this depends on the unknown 
strength of the feature.  One would like to include many lines to increase the S/N, but
this involves adding in lines of lower and lower column density.  At some point, 
the feature will become so weak that including more lines essentially adds only noise,
thereby reducing the significance of any detection.  As pointed out by \citet{lu91}, there
should therefore be some optimal lower limit which maximizes the S/N of the result.  To find
the optimal detection, we let the lower limit range from $10^{13.0}$ to $10^{13.9}$ \percm.
The resulting 78th percentile spectra are shown in Figure~\ref{fig:lspec} in the velocity
space of the \ion{O}{5} \lam 630 feature.  The number of lines and the \ion{O}{5} equivalent
width for each sample, measured as described in \S\ref{sec:measew}, are listed in 
Table~\ref{ta:ldata}.  One can clearly see how the noise is reduced by including more lines,
and the \ion{O}{5} feature becomes well-defined.  In none of the samples do we find any evidence
for \ion{O}{4} \lam 788.

In order to assess the errors on these results and their significance, we perform Monte Carlo
simulations in which we randomize the redshifts of the \lya\ features so that they
can take on any value in the allowable range, i.e., between \lyb\ and 5000 \kms\ blueward of
\lya\ for the particular object, as defined previously.  The resulting composites should
thus consist of random noise, and by measuring the equivalent widths in the simulated spectra
in the same way as for the real data, we can evaluate the errors on our equivalent width
measurements.  We have performed 1000 simulations for each sample, and the rms
deviations of the equivalent width in these Monte Carlo spectra are shown as the errors on the 
\ion{O}{5} equivalent width in
Table~\ref{ta:ldata}.  The mean equivalent width in these simulations is slightly less than zero, 
typically by $-0.5$ m\AA, indicating that there is a slight 
zero-point offset associated
with our measurement technique that is likely due to the continuum placement.
Because for our analysis we compare our results to simulated data, which also
include this effect, we do not add it into our measurement of the line equivalent width.

These simulations can also be used to answer the question of how likely it is that
the feature is due to random chance.  In column 4 of Table~\ref{ta:ldata}, we list the fraction
of simulated equivalent widths that were less than that measured in the real data, which
can be interpreted as the probability that the result is not due to chance.  For a lower
limit of $10^{13.5}$ \percm, none of the simulations exceeded the measured value, so we list
the probability as $>0.999$.  
The significance maximizes with a lower limit around $10^{13.2}$ to $10^{13.7}$ \percm.  Because so
few (two or fewer) of the simulations exceed the real measurements for lower limits of 
$10^{13.2}$, $10^{13.3}$, and $10^{13.5}$ \percm, we cannot reliably distinguish which is
of greater significance.  However, we find that the distribution of equivalent widths in the 
simulations are well-described by a Gaussian, and therefore one can simply use Gaussian 
statistics to estimate the significance. 
The best detections correspond to approximately a 3$\sigma$ fluctuation
above the mean noise level and are significant at greater than a 99\% 
confidence level.

\begin{tablehere}
\caption{\ion{O}{5} EWs for Weak \lya\ absorbers:  \nhi$< 10^{14.2}$ \percm\label{ta:ldata}}
\centerline{
\begin{tabular}{lccc} \hline \hline
$\log N_{\rm min}$ (\ion{H}{1})	& $n$ &
\ion{O}{5} EW (m\AA)	& $P_{\rm real}$\\
\hline
13.9	& 69	& $19.0\pm 8.9$	& 0.976\\
13.8	& 96	& $15.7\pm 7.4$ & 0.987\\
13.7	& 136	& $17.3\pm 6.2$	& 0.995\\
13.6	& 163	& $14.2\pm 5.5$	& 0.996\\
13.5	& 203	& $14.9\pm 5.0$	& $>0.999$\\
13.4	& 243	& $10.8\pm 4.8$	& 0.987\\
13.3	& 298	& $12.5\pm 4.1$	& 0.998\\
13.2	& 347	& $10.9\pm 3.7$	& 0.999\\
13.1	& 394	& $8.5\pm 3.7$	& 0.987\\
13.0	& 444	& $7.3\pm 3.3$	& 0.956\\
\hline
\end{tabular}}
\end{tablehere}
\vspace{0.4cm}

\begin{tablehere}
\caption{Effect of Percentile on \ion{O}{5} Feature Significance\label{ta:perc}}
\centerline{
\begin{tabular}{lcc} \hline \hline
Percentile	&
\ion{O}{5} EW (m\AA)	& $P_{\rm real}$\\
\hline
0	& $132.4\pm 148.9$	& 0.778\\
10	& $-0.4\pm 24.7$	& 0.568\\
20	& $7.0\pm 13.3$		& 0.820\\
30	& $12.5\pm 11.3$	& 0.904\\
40	& $13.0\pm 8.7$		& 0.950\\
50	& $11.1\pm 7.2$		& 0.957\\
60	& $10.6\pm 5.9$		& 0.970\\
70	& $11.7\pm 5.0$		& 0.995\\
80	& $9.7\pm 3.6$		& 0.997\\
90	& $7.9\pm 3.8$		& 0.986\\
100	& $14.6\pm 16.5$	& 0.813\\
\hline
\end{tabular}}
\end{tablehere}
\vspace{0.4cm}

Since the low-column-density samples contain many lines, we can use them to illustrate the
validity of our choice of taking the 78th percentile in forming the composite spectrum.  
As our test case we use
the \nhi $> 10^{13.2}$ \percm\ sample.  At the top of Figure~\ref{fig:perc}, we show
the 78th percentile spectrum with 347 lines in the sample.  Below this, we show various
other percentiles of the contributing spectra, from the 10th (lowest) to the 90th (highest),
plotted at true scale.  We can measure the equivalent width of the feature in each percentile
and calculate the noise with Monte Carlo simulations as before.  These results are shown
in Table~\ref{ta:perc}.  As expected, although the equivalent width of the feature is roughly
constant, the noise varies considerably.  The gain in S/N by using the 78th percentile over, for example,
the median is even greater than what would be expected based on our simulations in \S\ref{sec:combine}.
This is probably because the noise is due mostly to random absorption features which span several pixels
and thus cause correlated errors in the continuum.  We list the significance of the
detection in each percentile in Table~\ref{ta:perc}, and none of these exceed the significance
achieved with the 78th percentile, although the results are similar for the 80th percentile.
Based on the simulations in \S\ref{sec:combine} and
the success of the 78th percentile when applied to the real data, we feel justified in using
this technique.

\vspace{0.2cm}
\begin{figurehere}
\centerline{\psfig{file=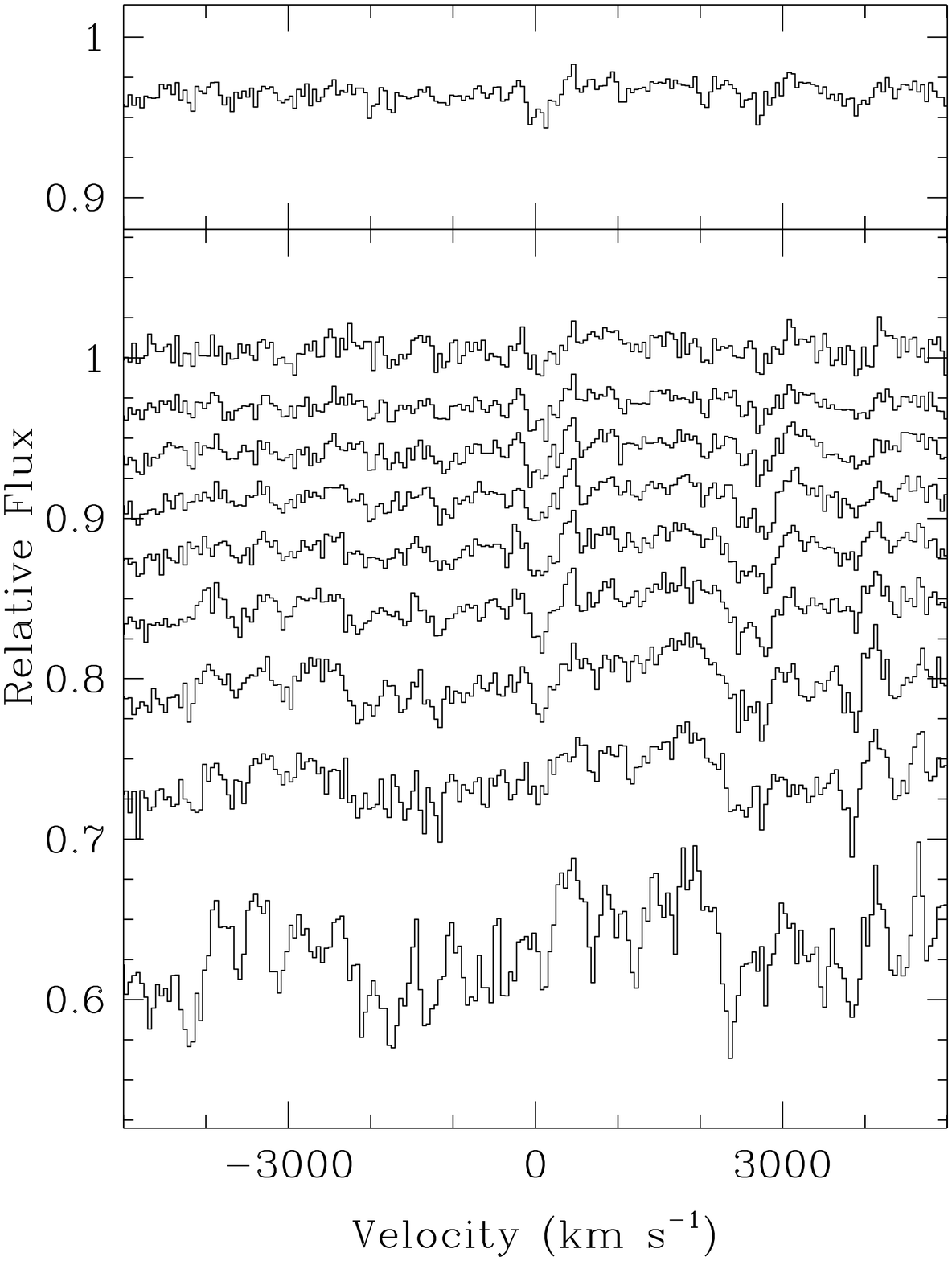,width=9cm}}
\caption{{\it Top}:  Composite 78th percentile spectrum of all lines with
$10^{13.2} <$\nhi $< 10^{14.2}$ \percm\ plotted in the velocity space of \ion{O}{5} \lam 630, 
as shown in Figure~\ref{fig:lspec}.
{\it Bottom}:  Various percentile spectra of the same data showing the effect of
the percentile on the result.  The spectra are in steps of 10 percent, ranging from the
10th percentile (bottom) to the 90th percentile (top).\label{fig:perc}}
\end{figurehere}
\vspace{0.6cm}

\subsection{Intermediate Column Densities:  $10^{14.2} <$\nhi $\lesssim 10^{15}$ \percm
\label{sec:res_i}}
\begin{figure*}
\centerline{\psfig{file=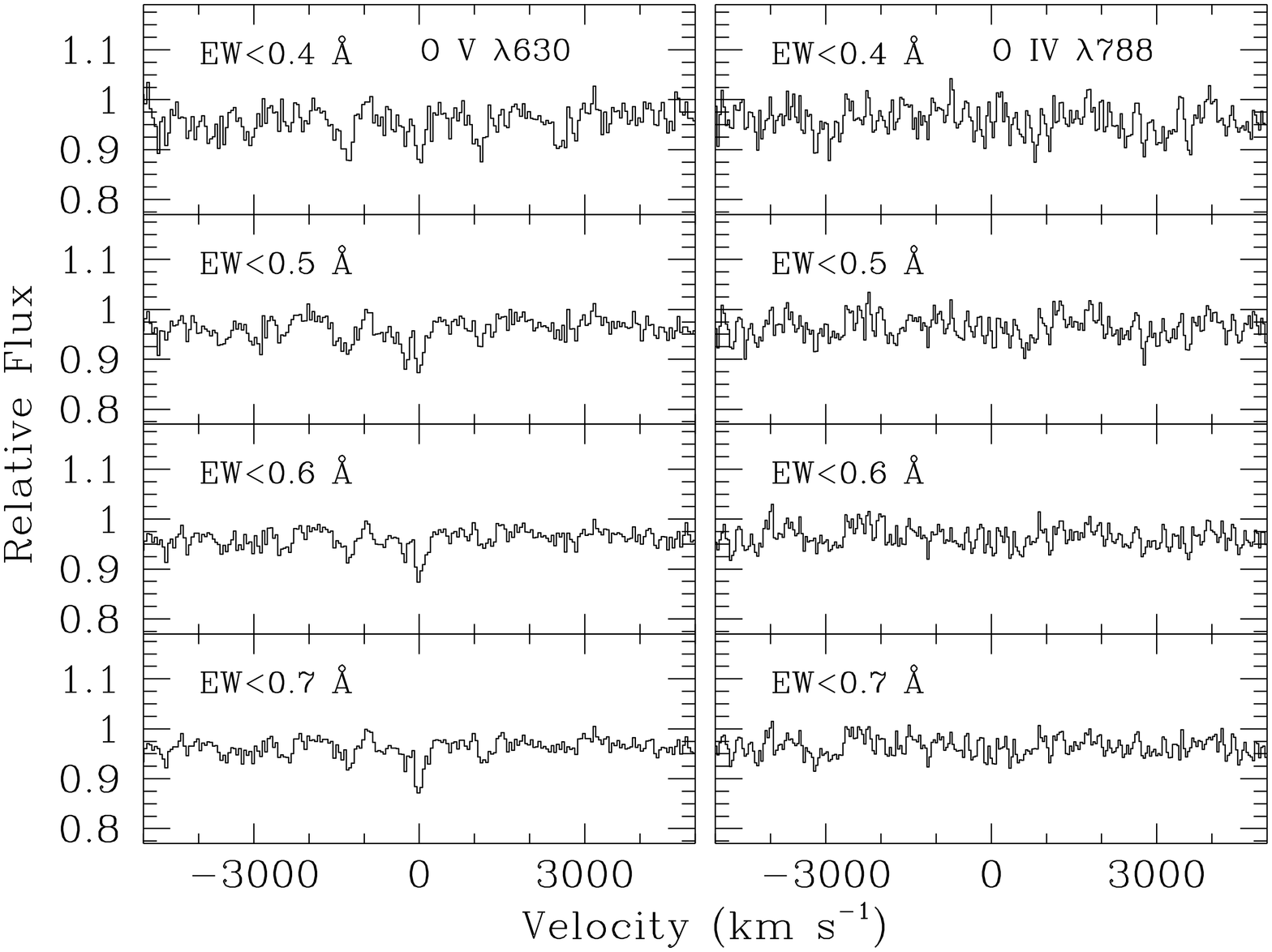, angle=-90, width=14cm}}
\caption{Composite spectra of \lya\ forest features with 
\nhi$> 10^{14.2}$ \percm\ for several maximum \lya\ equivalent widths.  The region
for \ion{O}{5} \lam 630 is plotted on the left, while the corresponding \ion{O}{4} \lam788 region
is shown on the right.  The \ion{O}{4} \lam788 feature is not detected in any of these spectra.
\label{fig:mspec}}
\end{figure*}

The division between the \lya\ forest and metal-line systems is not a precise concept.
In examining the intermediate-column-density lines we therefore try a range of upper limits 
on the \lya\ equivalent widths.  The measured equivalent widths of \ion{O}{5} 
are listed in Table~\ref{ta:mdata}.  The spectra are shown
in Fig~\ref{fig:mspec}, where we display not only \ion{O}{5} but also \ion{O}{4} \lam 788 to
illustrate that there is no evidence for a detection, which we quantify in 
\S\ref{sec:otherlimits}.
As opposed to the way we analyzed the low-column-density lines, here as we include more lines we 
are including lines of increasing strength and thus we definitely expect the significance of
the detection to increase.  This is clearly the case, and with a \lya\ EW upper limit of
0.6 \AA\ we can detect \ion{O}{5} with $> 4\sigma$ confidence.

\clearpage
\subsection{High Column Densities: \nhi $\gtrsim 10^{15}$ \percm \label{sec:res_h}}
For a $b$-parameter of 30 \kms, a \lya\ EW of 0.6 \AA\ 
corresponds to an \ion{H}{1} column density of $10^{16.0}$ \percm.  Since we excluded systems
with column densities $\gtrsim 10^{16.2}$ \percm\ based on Lyman-limit detections, lines with 
a \lya\ EW $\gtrsim 0.6$ \AA\ must be high-column-density systems with large $b$-parameters,
and we expect them to be associated with traditional ``metal-line systems''.
Although these strong systems are scarce and not the primary interest of this work, we can 
detect many absorption lines at such large column densities which 
further illustrates the success of our technique.

\begin{figure*}
\centerline{\epsfig{file=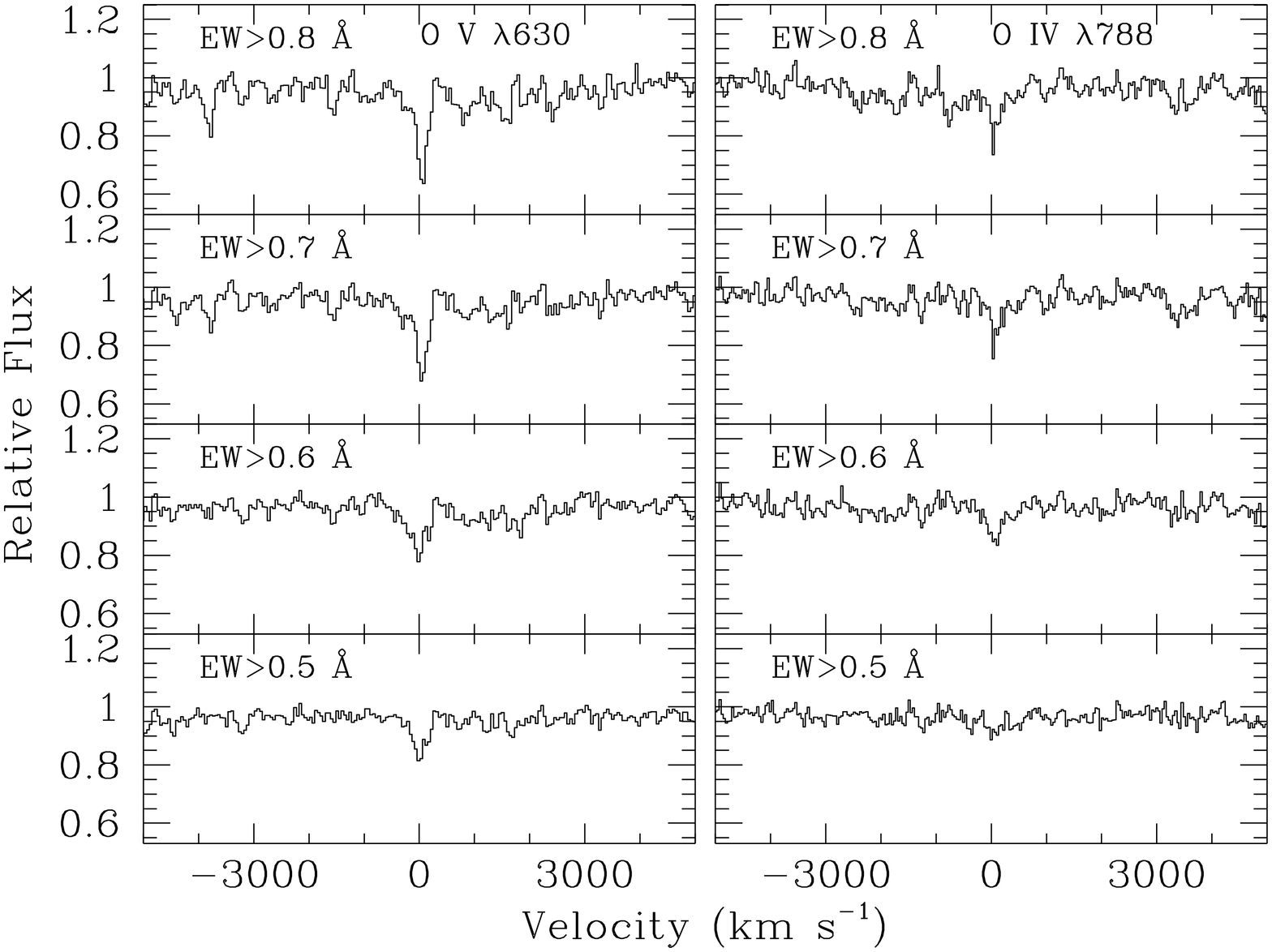, angle=-90, width=14cm}}
\caption{Composite spectra of the strongest \lya\ forest features in our sample 
for several minimum \lya\ equivalent widths.  \ion{O}{5} \lam630 is shown on the left, 
\ion{O}{4} \lam788 on the right.
\label{fig:hspec1}}
\end{figure*}

\begin{figure*}
\centerline{\epsfig{file=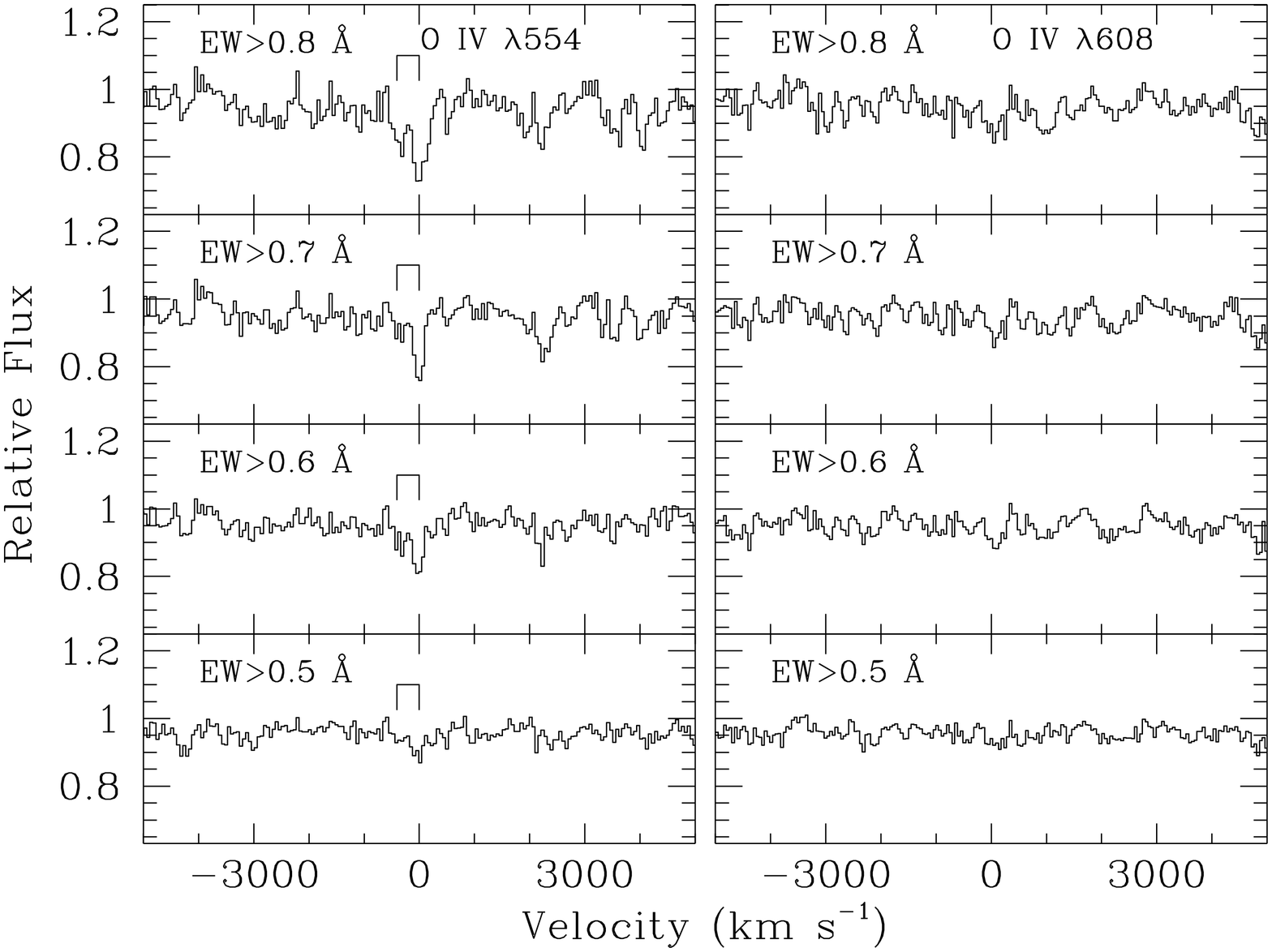,angle=-90,width=14cm}}
\caption{Same composites as Figure~\ref{fig:hspec1}, except \ion{O}{4} \lam554 is shown on the left
and \ion{O}{4} \lam608 on the right.  The \ion{O}{4} spectra are plotted in the velocity space of
the 554.07 \AA\ line.  Above the \ion{O}{4} spectra we indicate the positions of this line and the 
doublet companion at 553.33 \AA.
\label{fig:hspec2}}
\end{figure*}

\begin{figure*}
\centerline{\epsfig{file=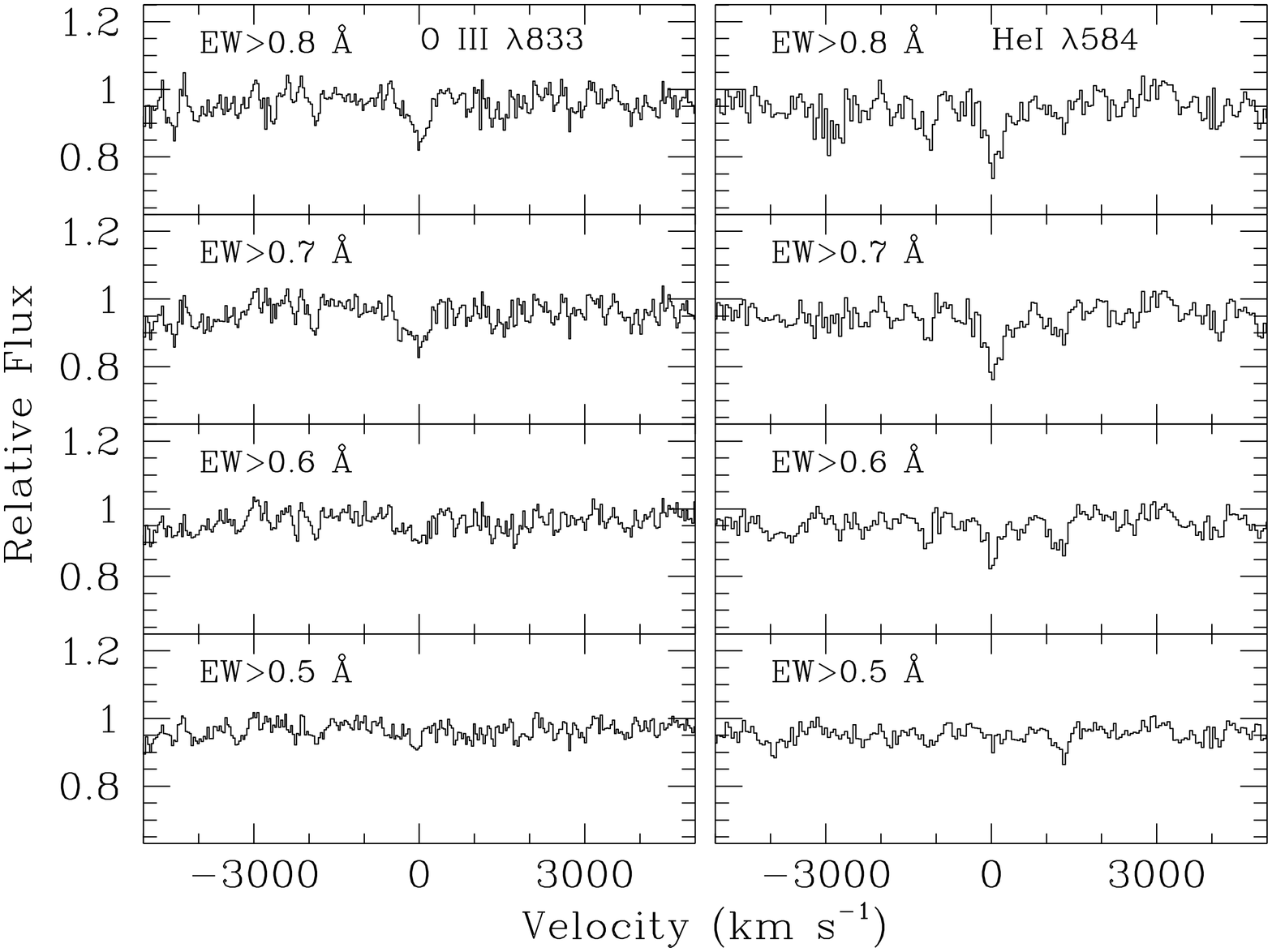,angle=-90,width=14cm}}
\caption{Same composites as Figures~\ref{fig:hspec1} and \ref{fig:hspec2} with \ion{O}{3} 
\lam833 shown on the left and \ion{He}{1} \lam584 on the right.
\label{fig:hspec3}}
\end{figure*}

In Figures~\ref{fig:hspec1}--\ref{fig:hspec3} we show plots for \ion{O}{5} \lam630, 
\ion{O}{4} \lam 788, \ion{O}{4} \lam 554, \ion{O}{4} \lam 608, \ion{O}{3} \lam 833, and
\ion{He}{1} \lam 584.  These lines have all been observed on an individual basis in Lyman-limit
systems \citep{revo93, vore93} and metal-line systems \citep{vore95}.  
The measured equivalent widths with errors from Monte Carlo simulations
are listed in Table~\ref{ta:hdata}.  The \ion{O}{4} line at 554.07 \AA\ has a weaker doublet 
companion at 553.33 \AA\ which is visible in Figure~\ref{fig:hspec2}.
We measure only the equivalent width of the \lam 554 line, although the presence of the
\lam 553 line probably affects this measurement slightly.  All of the lines are detected with
fairly high significance ($>0.98$) in the EW$>0.6$--0.8 \AA\ samples, with the exception
of \ion{O}{4} \lam 608.  By itself the presence of this line would be tenuous at best,
but given the strength of the other \ion{O}{4} lines it is present at about the expected level.
\vspace{0.7cm}
\begin{tablehere}
\caption{\ion{O}{5} EWs for Intermediate \lya\ Absorbers: $10^{14.2} <$ \nhi $\lesssim 10^{15}$ \percm
\label{ta:mdata}}
\centerline{
\begin{tabular}{lccc} \hline \hline
Max \lya\ EW (\AA)	& $n$ &
\ion{O}{5} EW (m\AA)	& $P_{\rm real}$\\
\hline
0.4	& 20	& $37.8\pm 19.1$	& 0.962\\
0.5	& 36	& $48.3\pm 13.9$	& 0.998\\
0.6	& 49	& $45.3\pm 11.4$	& $>0.999$\\
0.7	& 55	& $48.2\pm 10.5$	& $>0.999$\\
\hline
\end{tabular}}
\end{tablehere}
\vspace{0.7cm}

\begin{table*}
\caption{\ion{O}{5}, \ion{O}{4}, \ion{O}{3}, and \ion{He}{1} EWs for Strong \lya\ Absorbers:  
\nhi $\gtrsim 10^{15}$ \percm \label{ta:hdata}}
\centerline{
\begin{tabular}{lccccccc} \hline \hline
& & \multicolumn{6}{c}{Equivalent Width (m\AA)}\\
\cline{3-8}
Min \lya\ EW (\AA)	& $n$ &
\ion{O}{5} \lam630	& \ion{O}{4} \lam788 &
\ion{O}{4} \lam554	& \ion{O}{4} \lam 608 &
\ion{O}{3} \lam833	& \ion{He}{1} \lam 584\\
\hline
0.8	& 11	& $156.6\pm 25.6$ 	& $97.3\pm 39.2$	& $124.3\pm 34.6$ &
$52.3\pm 28.6$	& $120.9\pm 45.2$	& $110.3\pm 29.7$\\
0.7	& 14	& $146.1\pm 21.7$ 	& $91.2\pm 36.7$	& $68.6\pm 28.7$ &
$30.4\pm 25.2$	& $113.8\pm 40.4$	& $111.0\pm 25.1$\\
0.6	& 20	& $99.6\pm 18.4$ 	& $92.2\pm 27.7$	& $77.3\pm 22.7$ &
$32.9\pm 21.3$	& $62.7\pm 30.5$	& $64.0\pm 20.5$\\
0.5	& 33	& $85.7\pm 13.7$ 	& $44.8\pm 21.5$	& $49.2\pm 17.2$ &
$19.0\pm 15.6$	& $29.6\pm 22.9$	& $4.7\pm 15.4$\\
\hline
\end{tabular}}
\end{table*}

\subsection{Limits on Other Ions\label{sec:otherlimits}}
Using the same Monte Carlo simulations that we used to determine error bars
on the observed features, we can define 90\% confidence limits
on the equivalent widths of absorption for unobserved ions.
We do this not only for \ion{O}{4} \lam 788 absorption
but also for several other ions which have transitions which one might expect to observe in
the EUV.  Specifically, we constrain \ion{He}{1} \lam 584, \ion{O}{3} \lam 833, \ion{N}{4} \lam 765,
\ion{Ne}{5} \lam 568, \ion{Ne}{6} \lam 559, \ion{Ne}{8} \lam 770, and \ion{Mg}{10} \lam 610. 
We concentrate on three specific samples:  (1) the absorbers with 
$10^{13.2} <$\nhi $< 10^{14.2}$ \percm, which we refer to as sample L or the low-column-density
sample, (2) the absorbers with \nhi $> 10^{14.2}$ \percm\ and \lya\ equivalent widths
$< 0.6$ \AA, which we will to as sample I or the intermediate-column-density sample, and
(3) the absorbers with \lya\ equivalent widths $>0.6$ \AA, which we refer to as sample H or the 
high-column-density sample.
The equivalent width limits for various unobserved ions are listed in 
Table~\ref{ta:otherions} for all three samples.  For completeness we include the measured 
equivalent width for detected ions.  The ratios of the these ions to \ion{H}{1} are 
discussed in \S\ref{sec:discuss}.

\begin{table*}
\caption{Equivalent Widths and Column Density Ratios to \ion{H}{1}\label{ta:otherions}}
\centerline{
\begin{tabular}{lcccccc} \hline \hline
& & \multicolumn{2}{c}{Sample L} & \multicolumn{2}{c}{Sample I} & Sample H\\
\cline{3-4} \cline{5-6} \cline{7-7}
Ion	& $\lambda$ (\AA)	&
EW(m\AA)	& \xihi 	&
EW(m\AA)	& \xihi	& EW(m\AA)\\
\hline
\ion{He}{1}	& 584.33 & $< 6.0$ & $<-1.0$ & $<16.4$ & $<-1.5$ & $64.0 \pm 20.5$\\
\ion{N}{4}	& 765.14 & $<6.2$  & $<-1.5$ & $<23.3$ & $<-1.9$ & $< 37.0$\\
\ion{O}{3}	& 832.93 & $<8.1$  & $<-1.0$ & $<23.7$ & $<-1.4$ & $62.7 \pm 30.5$\\
\ion{O}{4}	& 787.71 & $<7.0$  & $<-1.0$ & $<18.8$ & $<-1.5$ & $92.2 \pm 27.7$\\
\ion{O}{5}	& 629.73 & $10.9\pm 3.7$ & $-1.6$ to $-0.6$ & $45.3\pm 11.4$ & $-1.7$ to $-0.8$ & $99.6 \pm 18.4$\\
\ion{Ne}{5}	& 568.42 & $<5.1$  & $<-0.7$ & $<16.6$ & $<-1.0$ & $<26.3$\\
\ion{Ne}{6}	& 558.59 & $<7.9$  & $<-0.7$ & $<20.1$ & $<-1.1$ & $<31.2$\\
\ion{Ne}{8}	& 770.40 & $<8.2$  & $<-0.8$ & $<25.3$ & $<-1.1$ & $<38.9$\\
\ion{Mg}{10}	& 609.79 & $<4.8$  & $<-0.6$ & $<17.2$ & $<-1.0$ & $<28.4$\\
\hline
\end{tabular}}
\end{table*}

\section{DISCUSSION\label{sec:discuss}}

\subsection{The \ion{O}{5} / \ion{H}{1} Ratio\label{sec:o5h1}}

The interpretation of composite spectra is not straightforward.
In order to reliably interpret the measured equivalent widths, it is necessary to generate
simulated data to compare to the results.  As a simplification, we first will
assume that the mean ratio \ovhi\ is constant within each of
our subsamples.
This may or may not be a good assumption, depending on the ionization model,
although our subsamples span a small enough range of \nhi\ that we do not expect 
\ovhi\ to vary by more that a factor of $\sim 2$.
Making this assumption allows us to derive results that are, as much as possible, model
independent.  Later in this section we will drop this restriction and simulate data using
particular ionization models.  Our Monte Carlo simulations proceed as follows:
\begin{enumerate}
\item{We randomize the redshift of each line to take on any value in the allowable range, as
we did for our noise simulations in \S\ref{sec:results}.}
\item{We calculate the \ion{O}{5} column density for each absorber from the 
\ion{H}{1} column density and the given relative abundance \ovhi.}
\item{The equivalent width of \ion{O}{5} \lam 630 is calculated from the \ion{O}{5} column
density assuming a particular $b$-parameter.}
\item{For the particular (randomized) redshift of each absorber, we insert a Gaussian absorption
feature with a FWHM of 290 \kms\ and the appropriate equivalent width at the position of the
\ion{O}{5} \lam 630 absorption.}
\item{The absorbers are combined into a composite and the equivalent width of \ion{O}{5}
\lam 630 is measured as we have done for the real data.}
\end{enumerate}

An important assumption in this process is the $b$-parameter used to calculate the
equivalent width.  For the weak lines in the low-column-density sample, this is
fairly unimportant, since most of the lines will be reasonably near the linear portion of the
curve of growth.  It will have some effect, though, and it is worth considering since this
will have a more pronounced effect when we model lines at higher column densities.  
Individual metal lines as observed in \ion{C}{4} absorption typically have quite small
$b$-parameters.  As examples, \citet{cskh95} find a median $b$ for \ion{C}{4} lines
of 10 \kms, while \citet{essp00} find a median $b$ of 13 \kms.  However, when the \ion{C}{4} 
absorption is strong, the absorption corresponding to particular \lya\ features often consists 
not of a single line but several such narrow absorption features.
The effective $b$-parameter for such metal-line
clusters, when viewed as a single absorption feature, is significantly larger than 10 \kms.
Because individual metal lines are not generally detected at the small \ion{H}{1} column densities
that we probe, we do not know the typical structure of the metal-line absorption. Strong \ion{C}{4}
absorbers often contain perhaps 3--5 components (see, for example, the \ion{C}{4} profiles in
\citet{elpc+99}).  An effective $b$-parameter of 50 \kms\ should therefore provide a reasonable 
extreme value for the typical $b$-parameter of the metal absorption associated 
with a sample of \ion{H}{1} lines.
To illustrate the effect of $b$ on our results, we create simulated data for $b$-parameters of 
10 and 50 \kms.  A physically realistic picture is probably somewhere in between, since it seems
likely that some absorbers will consist of several components whereas others may be dominated by
a single strong component.

The simulations also depend on the scatter in \ovhi\ at fixed \nhi, which depends
on the scatter in both the metal abundances and physical conditions of the absorbers.  
Using \ion{C}{4} absorption, \citet{rhs97} and \citet{hdhw+97} find a scatter of around an order of
magnitude, while the analysis of \citet{dhhk+98} suggests that the true scatter may be smaller as some
of the observed scatter could be attributable to fitting uncertainties.  
The individual systems measured by \citet{essp00} indicate an rms scatter of around
0.5 dex.  We generate simulated data with a Gaussian distribution in \ovhi\ with an rms scatter, 
which we denote as $\sigma_{\rm O}$, of 0.0, 0.5, and 1.0 dex.  The results of our
simulations of sample L for different values of $b$ and $\sigma_{\rm O}$ are plotted in 
Figure~\ref{fig:lowmodelo5},
compared with our observed value.  For each line in Figure~\ref{fig:lowmodelo5} we have
computed 100 simulated data sets for each point in steps of 0.1 in 
\ovhi, and the plotted value represents of the mean of these
simulations.  The $b$-parameter makes a small difference, but $\sigma_{\rm O}$ has a larger effect.
If we assume that $\sigma_{\rm O}$ is likely in the range 0.5--1.0 and allowing for the uncertainty 
in $b$, we get agreement with our observations for \ovhi$\approx -1.7$ to $-0.8$.

Generating simulated datasets for sample I is not as simple.
Because we do not have accurately measured column densities for these lines, we cannot simply
calculate the \ion{O}{5} column densities from the \ion{H}{1} column densities of the lines in
our sample.  Instead, we use simulated lists of random lines drawn from a column density
distribution with $\beta = -1.41$ for \nhi $< 10^{14.3}$ \percm\ and $\beta = -1.83$ for
$10^{14.3} <$ \nhi $< 10^{16.2}$ \percm, as shown in Figure~\ref{fig:cddf}.  We assume
a $b$-parameter distribution as derived by \citet{kity97}:  a Gaussian distribution in
$b$ with $\overline{b} = 23$ \kms\ and $\sigma_b = 14$ \kms, and a minimum $b$ given
by $b_{\rm min} = 14 + 4 \times \log [$\nhi$/ 12.5]$ \kms.  The distribution of $b$
is important because we have used a cut in equivalent width to define our sample, so we
must know the equivalent width of the lines in the simulated data.  In our full sample,
we have 69 \lya\ lines with $10^{14.2} <$\nhi $< 10^{16.2}$ \percm.  In Figure~\ref{fig:ew}
we plot a histogram of the equivalent width distribution of these lines, as well as the
average of 1000 simulated lists of \lya\ lines in this column density range.  The notable
difference between the two is the excess of large equivalent width lines, EW $\gtrsim 0.8$ \AA,
in the real data.  These are likely metal-line systems, as discussed earlier, though
they could also be blended lines.  The upper-limit
in \lya\ equivalent width of 0.6 \AA\ for sample I is thus chosen to help assure
that this sample is relatively free of metal-line systems.
At lower equivalent widths, it seems that the simulated data are
a good approximation of the real data.

In generating our simulated \ion{O}{5} data for sample I, we 
replace our sample absorbers with randomly generated line lists.  Given that we now have lines
with a realistic distribution of \ion{H}{1} column densities, we can calculate the 
\ion{O}{5} column densities and generate simulated data as we did for the lower column
densities.  Figure~\ref{fig:midmodelo5} shows the results of the simulations.  At these
larger \ion{O}{5} equivalent widths, $\sigma_{\rm O}$ makes less of a difference, but the results
are quite sensitive to the assumed $b$-parameter, since for a small $b$-parameter the
stronger lines will begin to saturate.  An effective $b$-parameter of
10 \kms\ is almost certainly unrealistically low for this sample, since \ion{C}{4} absorption 
in this regime often
shows structure, but it is useful to consider as an extreme case. Depending on the $b$-parameter, 
we find \ovhi\ should be in the
range $-1.6$ to $-0.6$ for sample I.  This agrees with what we derived for sample L, 
giving us confidence in our analysis and suggesting that \ovhi\ does not vary more than an order of
magnitude over the range of \ion{H}{1} column densities that we have considered, from $10^{13.2}$
to $\sim 10^{15}$ \percm.

\subsection{Other Ions\label{sec:otherions}}
To this point we have only analyzed \ion{O}{5} since it is the only ion we detect in samples
L and I.  However, in \S\ref{sec:otherlimits} we placed 90\% confidence limits
on the equivalent widths of absorption from other ions.
By creating model data as we have done for \ion{O}{5} \lam 630, we can convert these 90\%
confidence upper limits on the 
equivalent widths of metal lines of other ions into constraints on the column density ratio of the
corresponding ions to \ion{H}{1}.
For all of the ions we assume an rms scatter in \xihi\ of 0.5 dex for the simulated data.  
Table~\ref{ta:otherions} lists our derived values and limits on \xihi\ for samples L and I.  We do not
calculate values for sample H because our lack of knowledge of the \ion{H}{1} column densities makes
the results very uncertain.

\begin{figurehere}
\centerline{\psfig{file=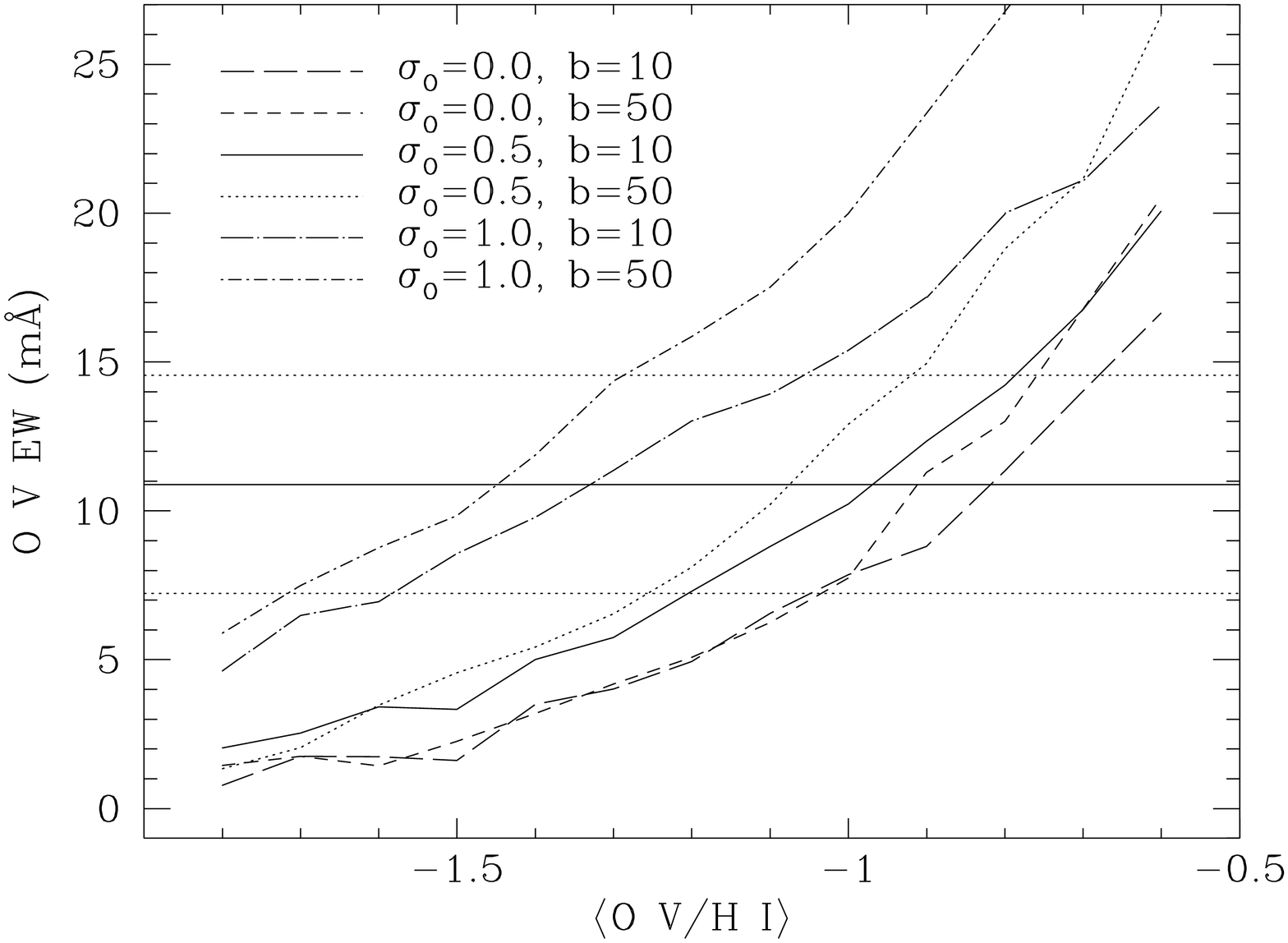,angle=-90,width=9cm}}
\caption{Expected \ion{O}{5} equivalent widths from simulated data for \lya\ absorbers with
$10^{13.2} <$ \nhi $< 10^{14.2}$ \percm, as a function of the mean ratio 
\ovhi.  The different curves correspond to different 
assumptions for the $b$-parameter for \ion{O}{5} and the rms scatter in 
\ovhi, denoted as $\sigma_{\rm O}$.  Our measured value is shown
as the horizontal solid line, and the $1\sigma$ limits on the measured value are shown as the 
horizontal dotted lines.
\label{fig:lowmodelo5}}
\end{figurehere}
\vspace{0.4cm}

\begin{figurehere}
\centerline{\psfig{file=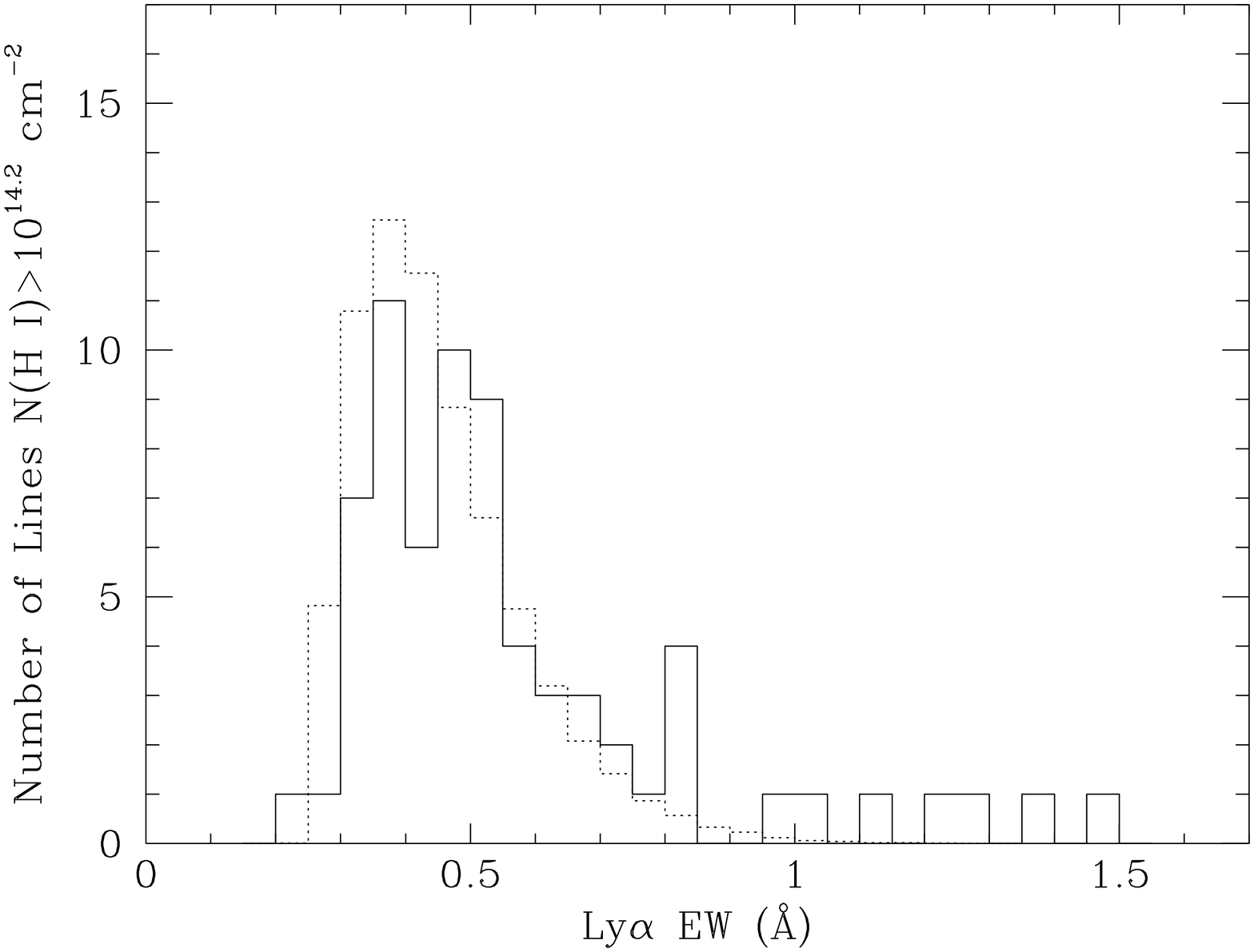,angle=-90,width=9cm}}
\caption{{\it Solid Line}:  Distribution of rest-frame equivalent widths of the 69 \lya\ 
absorption lines in our sample with \ion{H}{1} column densities $> 10^{14.2}$ \percm.  
{\it Dotted Line}:  Average distribution of rest-frame 
equivalent widths from 1000 randomly-generated lists of 69 \lya\ lines with column densities 
$> 10^{14.2}$ \percm, assuming the column density distribution shown in Figure~\ref{fig:cddf} 
and a $b$-parameter distribution as derived by \citet{kity97}.\label{fig:ew}}
\end{figurehere}
\vspace{0.4cm}

\subsection{The Oxygen Abundance\label{sec:abund}}

Converting a value for the \ovhi\ into an oxygen abundance
$\langle {\rm O} / {\rm H} \rangle$ requires knowledge of the relative ionization correction for
the ions \ion{O}{5} and \ion{H}{1}.  This is generally ascertained by means of 
photoionization models in which the IGM is illuminated by the processed radiation
of QSOs and/or stars.  We use the photoionization code CLOUDY \citep[version 94.00;][]{ferl96}.
Because we have not resolved individual \ion{O}{5} absorption components, in using these models to 
infer $\langle {\rm O} / {\rm H} \rangle$ we must necessarily assume that
the absorbers are predominantly single-phase; i.e., the \ion{O}{5} and \ion{H}{1} absorption occurs in
the same gas.
The main inputs to the models are the density and temperature of the absorbers as a function of
\ion{H}{1} column density, and the shape and normalization of the ionizing continuum.
For the physical conditions in the absorbers we use the results of \citet{hhkw98}, which are based
on hydrodynamical simulations as described by \citet*{kwh96}.  At $z=3$, \citet{hhkw98} find that 
the typical density of the absorbers in their model are well-described by a simple analytic fit,
$\log n_{\rm H} = -14.8 + 0.7 \log \,$\nhi\ assuming $\Omega_b h^2 = 0.0125$.  The typical 
temperature of the absorbers is described by $\log T=7.17 + 0.65 \log n_{\rm H}$ for 
$\log n_{\rm H} < -3.8$ and $\log T=3.18 - 0.4 \log n_{\rm H}$ for $\log n_{\rm H} > -3.8$.  The temperature
relation is somewhat dependent on the ionization history of the universe \citep{hugn97}, particularly
for underdense gas, but for the absorbers and redshifts we are studying this is a small effect and our
results are fairly insensitive to the temperature.

\begin{figurehere}
\centerline{\psfig{file=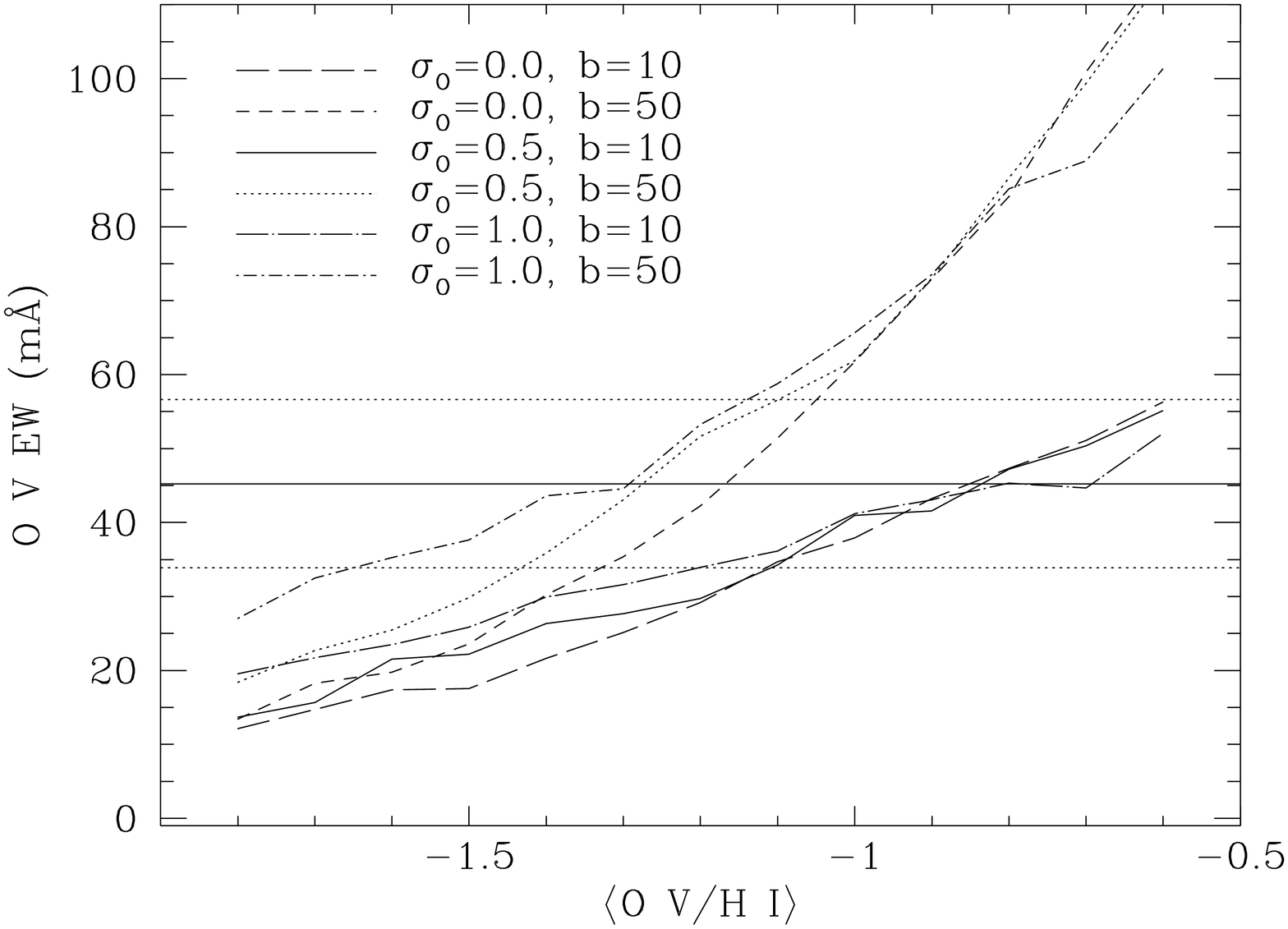,angle=-90,width=9cm}}
\caption{Same as Figure~\ref{fig:lowmodelo5} but for absorbers with 
\nhi $> 10^{14.2}$ \percm\ and a \lya\ equivalent width $< 0.6$ \AA.  In contrast to the weaker lines
in Figure~\ref{fig:lowmodelo5},
the scatter $\sigma_{\rm O}$ makes only a small difference but the result is more sensitive to the assumed
$b$-parameter, since many of the \ion{O}{5} features would begin to saturate if $b$ were small.
\label{fig:midmodelo5}}
\end{figurehere}
\vspace{0.4cm}

For our nominal ionizing spectrum we use the standard \citet[hereafter HM96]{hama96}
spectrum.  Specifically, we
use their result for $z=2$, although there is very little evolution in this spectrum between
$z\sim 2$ and 3.  \hm\ assume that QSOs are the source of the ionizing radiation for the IGM.
They assume a QSO spectral index of $\alpha_{\rm EUV}=-1.5$ ($f_\nu \propto \nu^\alpha$) for
the continuum shortward of the Lyman limit, consistent with the $\alpha_{\rm EUV}=-1.57\pm 0.17$
for radio-quiet QSOs derived by \citet{tzkd02}.  Some recent hydrodynamical simulations of the
IGM, however, suggest that a softer ionizing continuum than \hm\ is necessary to explain the
observed ratio of \ion{He}{2} to \ion{H}{1} in the IGM.  \citet{zanm97} require less
\ion{He}{2} ionizing radiation by a factor of $\sim 4$, while \citet{tlep+98} require a factor
of $\sim 2$ less.  \citet{cwkh97} resolve the discrepancy by increasing $\Omega_b h^2$ by an amount
consistent with recent determinations of $\Omega_b h^2$
from ${\rm D}/{\rm H}$ \citep{otks+01} and the cosmic microwave background \citep{dabb+02}.
Because the ions of interest to our discussion, in particular \ion{O}{5}, are
created by photons with wavelengths shortward of the \ion{He}{2} break, our results are sensitive
to differences in the \ion{He}{2} photoionization rate.  We create different ionizing spectra for our
photoionization models by first creating a simple analytic fit to the \hm\ spectrum, which we
find produces indistinguishable results for the ions of interest.  We then create softer spectra
by simply increasing the strength of the break at the \ion{He}{2} edge and then extending the 
softer spectrum to where it meets the \hm\ spectrum in the X-ray regime.  The original \hm\
spectrum and our model spectra are shown in Figure~\ref{fig:ionspec}.  For all these models we adopt
the normalization of the ionizing continuum of $J_\nu = 10^{-21.3}$ \jnu\ at 1 Ryd as derived by
\hm\ and in good agreement with \citet{gcdf+96}.

\begin{figurehere}
\centerline{\psfig{file=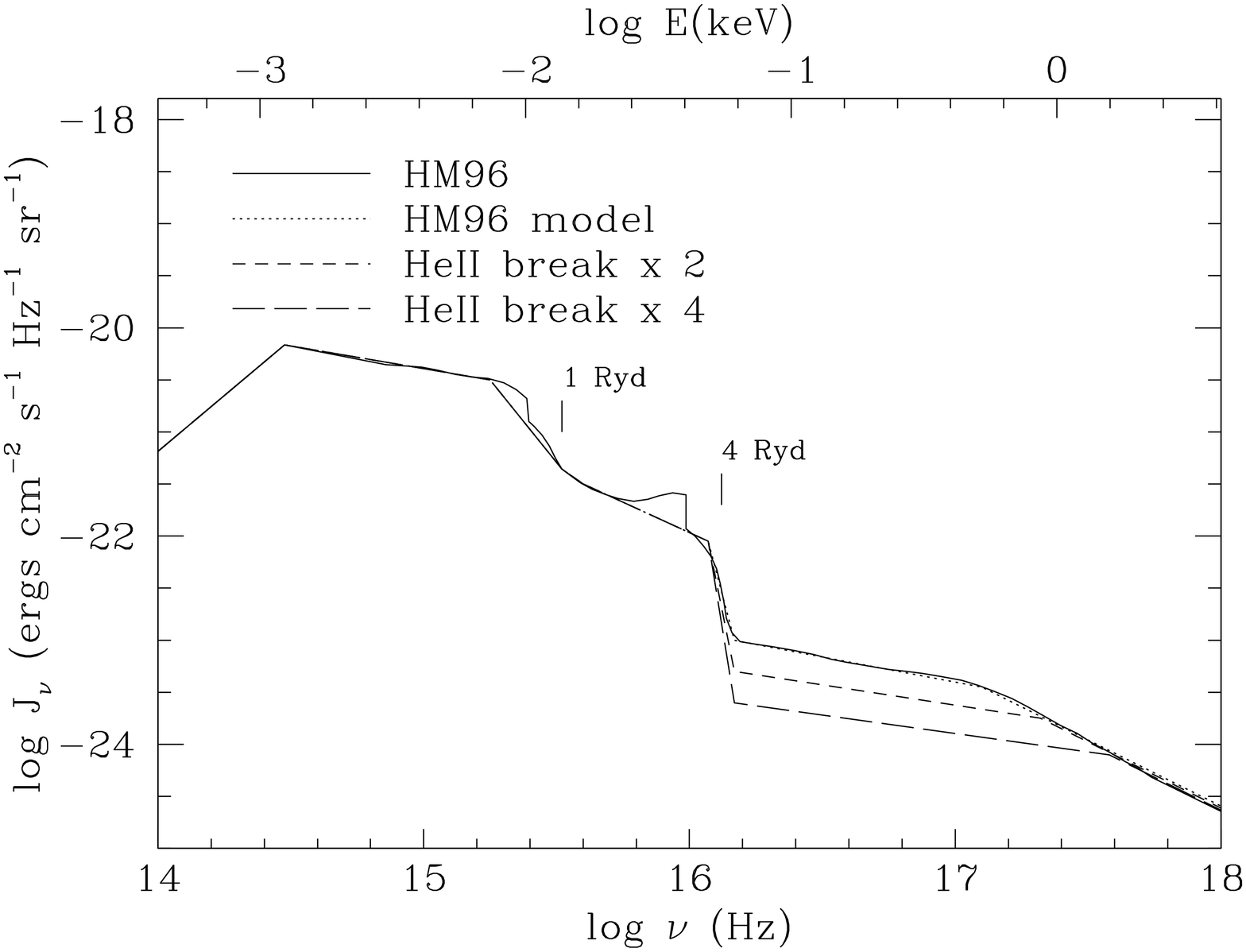, angle=-90, width=9cm}}
\caption{Various ionizing spectral shapes considered in this work.  The solid line shows the 
spectrum of \citet{hama96} at $z=2$.  The dotted line shows an approximation to this spectrum
using just a few points.  These two spectra yield virtually indistinguishable results for the ions
we consider.  The dashed lines show the same model spectrum, except we have artificially
increased the break at the \ion{He}{2} ionization edge by factors of two and four.
\label{fig:ionspec}}
\end{figurehere}
\vspace{0.4cm}

In Figure~\ref{fig:modelic} we plot the ionized fractions of \ion{O}{4}, \ion{O}{5},
\ion{O}{6}, and \ion{C}{4} relative to \ion{H}{1} for our photoionization models.  For the
\hm\ spectrum, \ovhi $-$\oh\ remains relatively constant for
$10^{13} <$\nhi $< 10^{14.8}$ \percm\ at around 3.6.  Given our values of \ovhi\ for sample L, this implies
\oh\ around $-5.3$ to $-4.4$, or $[{\rm O} / {\rm H}]$ around $-2.2$ to $-1.3$ with respect to the standard
solar oxygen abundance.  In the softer ionizing spectra models, slightly less oxygen is required to explain
our measurement for sample L but slightly more oxygen for sample I.
At larger \ion{H}{1} column densities, \ovhi\ varies more dramatically with \nhi\ and the
interpretation of our measured value of \ovhi\ over a range of \nhi\ is less obvious.

To include the effect of a changing ionization state with \nhi, we create simulations using
these specific ionization models to determine \nov\ for the simulated absorption where the free
parameter is now \oh\ rather than \ovhi.  We calculate model equivalent widths for \ion{O}{4}
\lam 788 as well to compare with our limits since this can provide an interesting constraint on the 
ionization state.  In Figure~\ref{fig:lowmodelo} we plot the equivalent widths of \ion{O}{5} and
\ion{O}{4} as a function of \oh\ for sample L,
assuming $b=50$ \kms\ and $\sigma_{\rm O} = 0.5$ dex.  We do not plot the results for different 
values of the $b$-parameter and $\sigma_{\rm O}$, 
but the effects are very similar to those seen in Figure~\ref{fig:lowmodelo5};
i.e., the required \ovhi\ is more by around 0.1 for $b=10$ \kms, and less by around 0.4 for 
$\sigma_{\rm O} = 1.0$ dex.
As expected, the required \oh\ agrees with what we estimated above, with slightly less oxygen required
for the softer spectra.  For none of these ionizing spectra does the required amount of
oxygen imply that there should be a clearly observable feature for \ion{O}{4}, although for
the \ion{He}{2} break $\times$ 4 spectrum the expected equivalent width of \ion{O}{4} \lam 788
is around our 90\% confidence limit.
This comparison of \ion{O}{5} to \ion{O}{4} is virtually independent of our assumptions about 
the $b$-parameter and $\sigma_{\rm O}$ since the \ion{O}{5} and \ion{O}{4} equivalent widths essentially
scale by the same factor for changes in these parameters.

\begin{figurehere}
\centerline{\psfig{file=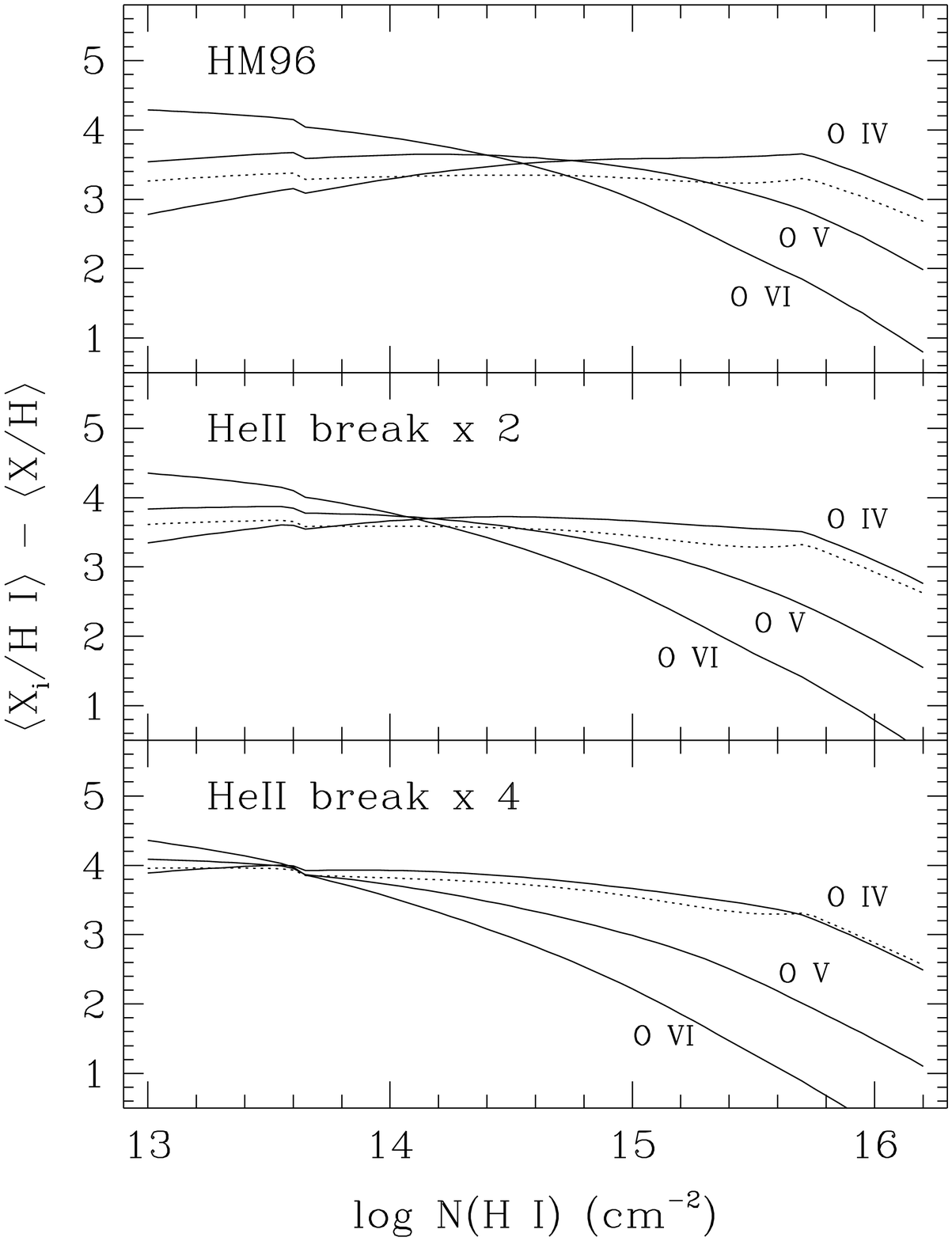,width=9cm}}
\caption{Ionized fractions of various ions relative to \ion{H}{1}.  The solid lines are for
\ion{O}{4}--\ion{O}{6} and are labeled in the diagram.  The dotted line is for \ion{C}{4}.
The models correspond to the ionizing spectra shown in Figure~\ref{fig:ionspec}.  A
normalization of $J_{\nu} = 10^{-21.3}$ \jnu\ at 1 Ryd
is used for all three models.
\label{fig:modelic}}
\end{figurehere}
\vspace{0.4cm}

The \ion{O}{4} limit is more interesting for sample I.
The predicted equivalent widths are plotted in Figure~\ref{fig:midmodelo_50} assuming 
$\sigma_{\rm O} = 0.5$ dex and $b=50$ \kms,
which as we argued before is probably more realistic than $b=10$ \kms\ for \nhi $> 10^{14.2}$ \percm.
Again, the effects of changing the $b$-parameter and $\sigma_{\rm O}$ are similar to those in
Figure~\ref{fig:midmodelo5}.
For these absorbers, 
the simulated data suggest that the observed lack of \ion{O}{4} favors a hard ionizing spectrum.
For the \hm\ spectrum the predicted equivalent width is only slightly higher than our 90\%
confidence limit.
However, for the \ion{He}{2} break $\times$ 4 spectrum, the simulations suggest an equivalent width of
\ion{O}{4} of more than three times our limit.

This constraint is sensitive to the normalization of the ionizing continuum.
An increase in the ionizing flux will decrease \oivov\ resulting in less predicted \ion{O}{4}.
We rerun the simulations using $J_\nu = 10^{-21}$ \jnu\ at 1 Ryd  as found by \citet*{cec97}.
The resulting \oivov\ is smaller by around 30--40\%, 
consistent with the \hm\ spectrum but still lying above our upper limit for the softer spectra.
Density gradients in the absorbers can also influence the results.  By using a constant density,
we are assuming that the absorption takes place primarily in the densest portion of the absorber.
This should be a good approximation for a given ion provided that the volume density of the ion
increases with total density.  However, sample I contains absorbers with \ion{H}{1} densities
near the point where \ion{O}{5} peaks.  If a significant amount of the absorption were to occur
in the density wings of the absorber, the observed \oivov\ would be less than what we have 
calculated for a constant-density model.

\begin{figurehere}
\centerline{\psfig{file=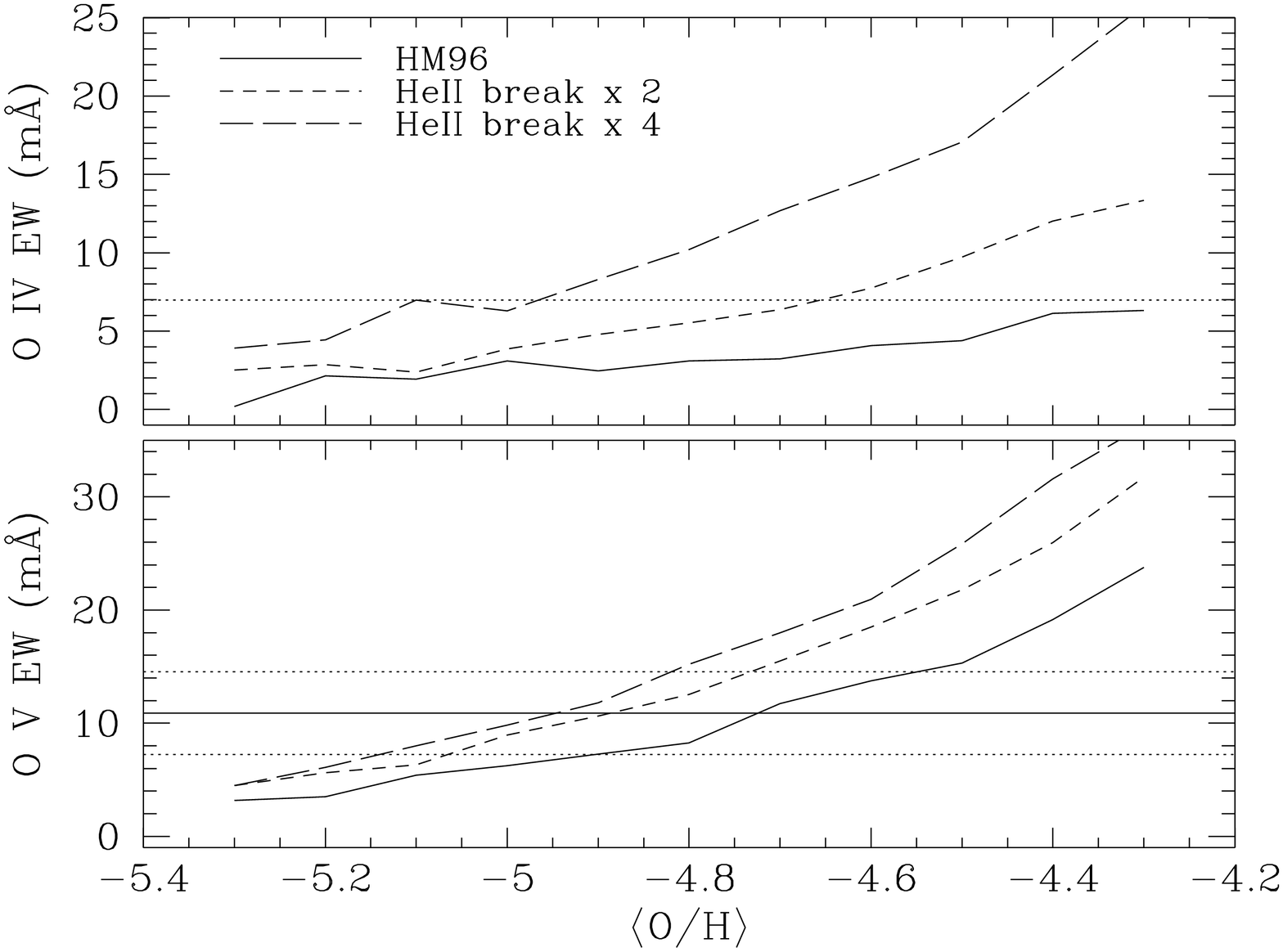,angle=-90,width=9cm}}
\caption{Expected equivalent widths of \ion{O}{5} \lam 630 and \ion{O}{4} \lam 788
as a function of \oh\ for \lya\ absorbers with
$10^{13.2} <$ \nhi $< 10^{14.2}$ \percm.  The different lines correspond to different
assumed ionization fractions, as shown in Figure~\ref{fig:modelic}, based on different
ionizing spectra.  For all the simulated data we assume an rms scatter in 
\oh\ of 0.5 dex and a $b$-parameter of 50 \kms.  Our 90\%
confidence upper limit on the equivalent width of \ion{O}{4} \lam 788 is shown as the dotted line
in the top panel.  Our measured value for \ion{O}{5} is shown in the bottom panel as it was
in Figure~\ref{fig:lowmodelo5}.
\label{fig:lowmodelo}}
\end{figurehere}
\vspace{0.4cm}

\begin{figurehere}
\centerline{\psfig{file=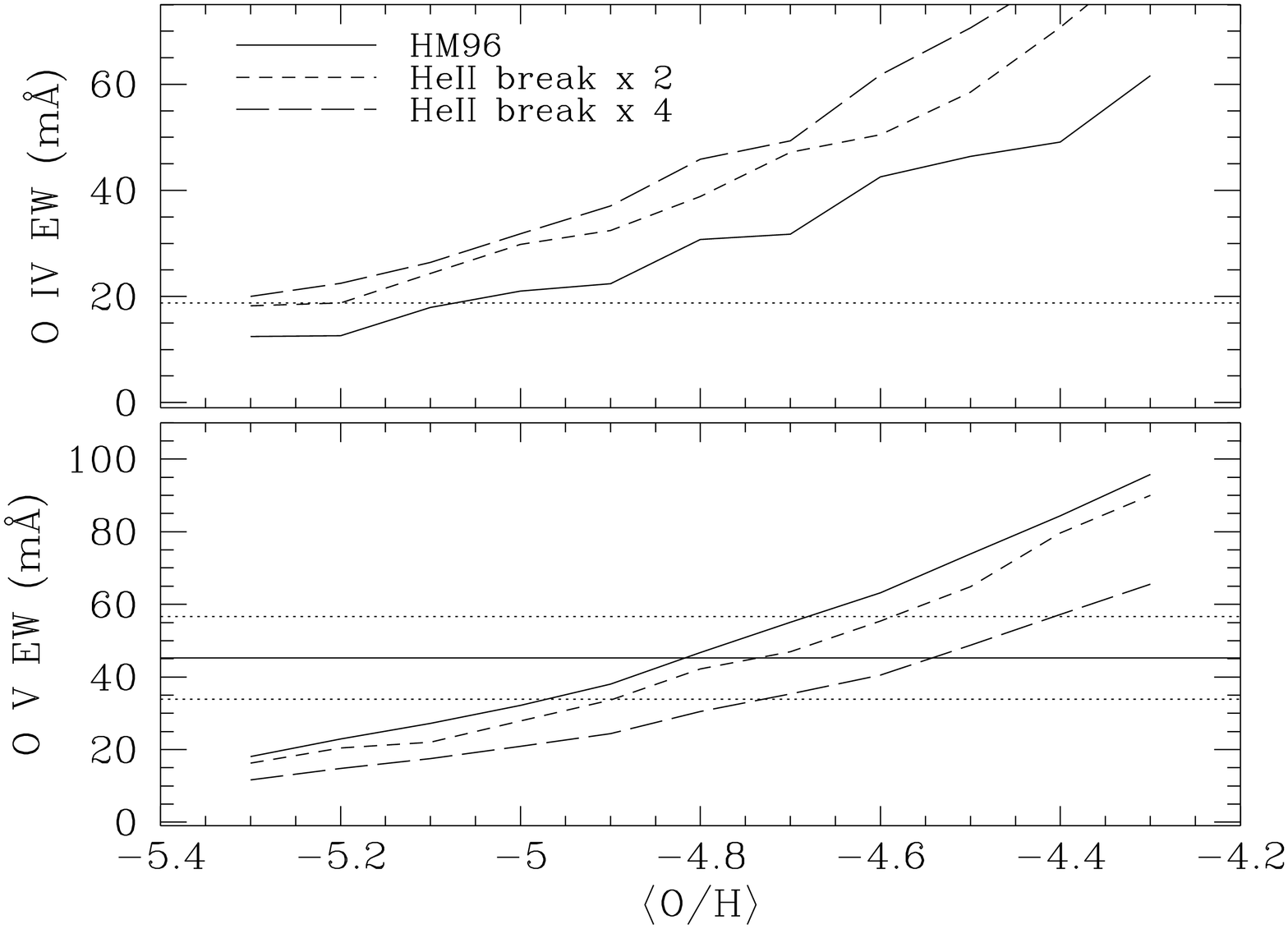,angle=-90,width=9cm}}
\caption{Same as Figure~\ref{fig:lowmodelo} for absorbers with \nhi $> 10^{14.2}$ \percm\ 
and a \lya\ equivalent width $< 0.6$ \AA.
\label{fig:midmodelo_50}}
\end{figurehere}
\vspace{0.4cm}

An additional consideration is our ignorance of the column densities of the lines in sample I.
Because there are only 49 lines in the sample, it is possible that the true distribution of \ion{H}{1}
column densities is significantly different than our simulated lists.  In particular, there could be
a relative lack in the true sample of high-column-density systems where \oivov\ is highest.  Our 
simulations suggest that the uncertainty in the predicted \ion{O}{4} equivalent width due to
statistical fluctuations in the column-density distribution is comparable to the random noise.
The additional uncertainly for \ion{O}{5} is much smaller since in our photoionization models
\nov\ varies much less with \nhi\ over the range of \ion{H}{1} column densities spanned by sample I.
Models with predicted \ion{O}{4} equivalent widths only moderately above our stated limit are
therefore consistent with our results.  However, it is quite difficult to 
reconcile our limit on \ion{O}{4} with the \ion{He}{2} break $\times$ 4 spectrum.  We
conclude that our data are probably inconsistent with such a soft ionizing spectrum and that the true
ionizing background is not significantly softer than that of \hm.

As we noted above, early observational constraints on $N($\ion{He}{2}$)/$\nhi, often called $\eta$,  
led us to consider ionizing continua softer than \hm.  The strongest 
constraint on $\eta$ generally considered in the literature is from the \ion{He}{2} opacity
measurement of
\citet*{dkz96}.  Recently, \citet{ksoz+01} analyzed a high-resolution spectrum of the \ion{He}{2}
absorbing region with the Far Ultraviolet Spectroscopic Explorer enabling for the first time the
explicit measurement of $\eta$ for individual absorbers.  Using both measured values of $\eta$
and lower limits for \ion{He}{2} absorbers with no corresponding \ion{H}{1} absorption, they
find a mean $\eta$ of $\sim 80$.  However, the mean $\eta$ for absorbers
with detectable \ion{H}{1} (\nhi $\gtrsim 10^{12.3}$ \percm) is $\sim 30$.  Because our sample
is selected by \ion{H}{1} absorption, it is appropriate to compare to this value.  
Comparing to the models of \citet*{fgs98}, $\eta = 30$ suggests a QSO source spectrum with
$\alpha_{\rm EUV}$ around $-1.6$, consistent with \citet{tzkd02} and similar to \hm.
Thus the fact that our data support a hard ionizing continuum is consistent with the \ion{He}{2}
results if one takes into account that an \ion{H}{1}-selected sample preferentially selects
absorbers with lower values of $\eta$ which are likely photoionized by hard,
QSO-like radiation.

Including uncertainty in the scatter, our models suggest \oh\ of around $-5.2$ to $-4.6$ for sample I,
although the required \oh\ could be much higher if the effective $b$-parameter is much smaller than
50 \kms.  This agrees with the results for sample L, for which the results are more robust
due to the fact that the \ion{H}{1} column densities of the constituent absorbers are directly measured.
We therefore conclude that the typical \oh\ for the IGM likely lies in the range $-5.3$ to $-4.4$, as
derived for sample L.  The abundance with respect to the standard solar value is thus 
$[{\rm O}/{\rm H}]$ around $-2.2$ to $-1.3$.

Having derived this abundance, it is interesting to 
compare this to results from studies of \ion{C}{4}.  Using the optical depth ratio technique,
\citet{essp00} find that \civhi $=-2.6$ for $\tau_{\rm Ly\alpha} \gtrsim 1$ provides a good match
to their data, in agreement with what \citet{soco96} found for individually detectable \ion{C}{4} 
absorbers.  Using our \hm\ model for the ionization correction for \ion{C}{4}, \civhi $=-2.6$
corresponds to $[{\rm C}/{\rm H}] \approx -2.5$.  This is in good agreement with what other authors
have derived assuming a \hm\ spectrum \citep{hdhw+97,dhhk+98}.  The inferred $[{\rm O}/{\rm C}]$
is in the range 0.3--1.2, assuming that the chemical composition does not change dramatically 
from $z\sim 2$ to 3.5 since our results are for slightly lower redshift than those for
\ion{C}{4}.  Similar results are obtained for the \ion{He}{2} break $\times$ 2
spectrum. This model implies less carbon, $[{\rm C}/{\rm H}] \approx -2.7$, to match the \ion{C}{4} 
results, but also less oxygen to match our results, resulting in the same $[{\rm O}/{\rm C}]$.
Thus, since the relative ionization corrections in our models for \ion{C}{4} and \ion{O}{5} 
vary little for the \ion{H}{1} column densities of interest, the uncertainty in $[{\rm O}/{\rm C}]$ is
dominated by our uncertainty in \ovhi.

An overabundance of oxygen relative to carbon is expected if the IGM is enriched in metals
by Type II supernovae.  \citet{eagl+93} find overabundances of oxygen relative to carbon by factors
of 3--5 in halo stars, consistent with the low end of the range we infer.  There is also
evidence that this overabundance may increase with decreasing metallicity \citep*{igr98}.
A relative overabundance of oxygen is expected from
Type II supernova yields \citep{wowe95}.  However, large values of $[{\rm O}/{\rm C}]$
may require nucleosynthesis via pair instability supernovae \citep{hewo02} from very massive
stars ($m \gtrsim 100 \ M_\odot$) in the early universe \citep{adsl+01,scha02}.
A bimodal initial mass function for Populations III stars, including a peak at $\sim 100 \ M_\odot$,  
is predicted by the hydrodynamical simulations of \citet{naum01}.  An additional 
predicted consequence of nucleosynthesis by very massive stars is a large abundance of silicon, 
$[{\rm Si} / {\rm C}] \sim 1.0$.  In contrast, \citet{soco96} infer a typical value of 
$[{\rm Si} / {\rm C}] \sim 0.4$ for metal-line systems from the observed \ion{Si}{4} / \ion{C}{4} ratios.
There is evidence for absorbers
with $[{\rm Si} / {\rm C}] \gtrsim 1.0$ at high redshift ($z>3$), but the assumed implausibility of such abundance ratios are used to argue instead for a substantial softening of the ionizing background 
\citep{scdf+97,gish97,song98}.  Further exploration of the ionizing background at $z>3$, such as from
measurements of the \ion{He}{2} / \ion{H}{1} and \ion{O}{4} / \ion{O}{5} ratios to be made possible by the 
Cosmic Origins Spectrograph, will help to clarify this issue.

Evidence for an overabundance of oxygen at high redshift 
has been found by other authors.  \citet{rvhe+92} find
$[{\rm O}/{\rm C}] \gtrsim 0.6$ for Lyman-limit systems.  Using the optical depth technique,
\citet{dhhk+98} find a good fit
for their data using $[{\rm O}/{\rm C}] = 0.5$ for \ion{C}{4} absorbers at $z\gtrsim 3$.  
However, based on the lack of narrow \ion{O}{6} absorption lines in their data, they claim that
the overall metallicity must drop dramatically, by a factor of at least $\sim 3$, for
absorbers with \nhi $\lesssim 10^{14.5}$ \percm.  
In contrast, our results suggest that there
is a large abundance of metals at \nhi\ $\lesssim 10^{14.2}$ \percm, consistent with what is 
found at higher \ion{H}{1} column densities.  The \ion{O}{6} detection of \citet{srsk00}
supports this result by suggesting that the enrichment of the IGM does indeed extend to 
densities below the universal mean.

It is instructive to compare our results to
\citet{srsk00}.  Assuming \ovhi$=-1.1$ and using the \hm\ model for the ionization 
correction, for \lya\ optical depths ranging from 0.1 to 1.0, corresponding to 
$10^{13.6} \lesssim$ \nhi $\lesssim 10^{14.6}$ \percm\ for $b=30$ \kms, we would predict
\ovihi\ from around $-0.5$ for $\tau_{\rm Ly \alpha} = 0.1$ to $-1.3$ for 
$\tau_{\rm Ly \alpha} = 1.0$.  The corresponding ratio of the optical depth of \ion{O}{6}
\lam 1032 absorption to \lya\ would range from $-1.1$ to $-1.9$.  This agrees well with the
apparent optical depths of \ion{O}{6} measured by \citet{srsk00} for the same redshift range,
though as they point out, the true optical depth of \ion{O}{6} could be considerably different
from the apparent optical depth due to the effects of noise and Lyman-line contamination.

The existence of metals at low column densities addresses the issue of the source of the metals
in the IGM.  The enrichment of the IGM is thought to occur in one of two basic ways:  (1)
by a high-redshift era of small star-forming regions, either Population III stars \citep{osgn96} or
protogalaxies \citep*{mfr01}, or (2) by in-situ star formation,
either in the \lya\ absorbers themselves or nearby galaxies.  The primary distinction between
the two scenarios is the volume extent of the enrichment.  While early supernovae could
be expected to pre-enrich the universe in a fairly uniform way, it is difficult to enrich the
most diffuse regions of the IGM by in-situ star formation.
For example, \citet{gnos97} predict
a strong dropoff in metallicity at column densities below around $10^{13.5}$--$10^{14.5}$ \percm\
at $z=3$.  Although our sample L includes absorbers down to $10^{13.2}$ \percm, we cannot 
be sure that there are metals at such low column densities, though the fact that the
S/N of the detection does not decrease until we include absorbers with \nhi $<10^{13.2}$ \percm\
certainly provides evidence that there are metals at low column density.  Our results apply to slightly
lower redshifts than this prediction, so there is more time for metals to diffuse into less dense
regions, and the absorbers that we observe correspond to slightly larger overdensities for the same
\nhi.  Given these considerations, we cannot conclusively claim that our results are inconsistent
with in-situ enrichment scenarios, which are currently an active area of study 
(\citealt*{fps00}; \citealt{ahsk+01}).
However, it is fair to say that our detection of \ion{O}{5} in sample 
L, in conjuction with previous detections of \ion{C}{4} and \ion{O}{6} in similar density regimes,
provide evidence for the presence of metals at quite low column densities, which would favor
a uniform enrichment as provided by an early era of star formation.

\section{SUMMARY\label{sec:summary}}
We have used {\it HST} FOS spectra of four QSOs to search for absorption features associated with
\lya\ forest absorbers in the EUV in the redshift range $1.6 < z < 2.9$.  
The results of this search can be summed up as follows:
\begin{enumerate}
\item{We detect \ion{O}{5} \lam 630 over a large range of 
\nhi.  Most interestingly, we detect \ion{O}{5} in a sample of absorbers with 
$10^{13.2} <$ \nhi\ $< 10^{14.2}$ \percm\ with greater than 99\% confidence.}
\item{We detect \ion{O}{4} \lam 788, \ion{O}{4} \lam 554, \ion{O}{3} \lam 833, and \ion{He}{1}
\lam 584 only for the strongest absorbers in our sample, those with \lya\ equivalent widths
$\gtrsim 0.6$ \AA.}
\item{We find no evidence for \ion{N}{4} \lam 765, \ion{Ne}{5} \lam 568, \ion{Ne}{6} \lam 559, 
\ion{Ne}{8} \lamlam 770, 780, or \ion{Mg}{10} \lamlam 610, 625 absorption in any of our samples.}
\end{enumerate}

For absorbers with \lya\ equivalent widths $\lesssim 0.6$ \AA, the \ion{O}{5} detections imply
\ovhi $\approx -1.7$ to $-0.6$ based on our simulated data, where this range allows for uncertainties
in the assumed $b$-parameter and scatter in \ovhi.  This result implies that \ion{O}{5} is typically
a factor of $\sim 10-100$ more abundant than \ion{C}{4}.  The \ion{O}{5} \lam 630 line thus has
EW$/\lambda$ a factor of $\sim 30-300$ larger than \ion{C}{4} \lam 1548, or ranging in strength from around
that of Ly$\delta$ to \lyb, making it an excellent tracer of metal content.  

Using photoionization models to calculate the ionization correction, we find that the oxygen 
abundance in the IGM is $[{\rm O}/{\rm H}] \approx -2.2$ to $-1.3$, implying 
$[{\rm O}/{\rm C}]\approx 0.3$ to $1.2$, consistent with what is found in halo stars and 
Lyman-limit systems.
The overabundance of oxygen suggests Type II supernova enrichment, but an unusual stellar initial
mass function resulting in a 
significant contribution from pair instability supernovae of 
very massive ($m \gtrsim 100 \ M_\odot$) Population III stars may be necessary.

The fact that we find no evidence for \ion{O}{4} \lam 788 absorption except in the strongest 
systems provides an interesting constraint on the ionizing background spectrum, specifically that
it is unlikely to be more than a factor of $\sim 2$ softer than the \hm\ spectrum at 4 Ryd.  A hard
ionizing spectrum is consistent with measurements of \ion{He}{2} absorption for absorbers with
detectable \ion{H}{1}.

We conclude that studying the absorption of the IGM in the rest-frame EUV is not only a useful
but essential tool for gaining a more complete understanding of the metal content and ionization
of the IGM.  From longer wavelength data alone one cannot place useful constraints on absorption
from multiple ionization stages of the same element in the diffuse IGM.  Ultimately, a more
complete understanding requires such analysis, and the clearest path to this end is through
the simultaneous study of multiple ionization stages of oxygen 
(\ion{O}{3}--\ion{O}{6}).  We look forward 
to the installment of the Cosmic Origins Spectrograph on the {\it HST}, currently planned for
2004, which will enormously increase our ability 
to observe weak oxygen features on an individual basis.

\acknowledgements
We are grateful to A. Songaila for providing us with the Keck spectrum of HE~2347--4342.

\end{document}